\documentclass[twoside,a4paper]{article}
\usepackage{amsmath,amssymb,amsfonts,qic}
\usepackage{epsfig}

\renewcommand{\thefootnote}{\fnsymbol{footnote}}

\newcommand{\Tr}{\operatorname{Tr}}
\newcommand{\rmd}{\mathrm{d}}
\newcommand{\rmi}{\mathrm{i}}
\newcommand{\rme}{\mathrm{e}}
\newcommand{\rmq}{\mathrm{q}}
\newcommand{\rmc}{\mathrm{c}}
\newcommand{\rmf}{\mathrm{f}}

\newcommand{\openone}{\leavevmode\hbox{\small1\normalsize\kern-.33em1}}

\begin{document}
\setlength{\textheight}{8.0truein}

\runninghead{Quantum measurements and entropic bounds}
            {A. Barchielli, G. Lupieri}

\normalsize\textlineskip
\thispagestyle{empty}
\setcounter{page}{1}

%\copyrightheading{Vol.}{No.}{Year}{Page Nos.}
%\copyrightheading{0}{0}{2005}{000--000}

\vspace*{0.88truein}

\alphfootnote

\fpage{1}

\centerline{\bf QUANTUM MEASUREMENTS AND ENTROPIC BOUNDS }

\vspace*{0.035truein}

\centerline{\bf ON INFORMATION TRANSMISSION}

\vspace*{0.37truein}

\centerline{\footnotesize Alberto Barchielli\footnote{Also: Istituto Nazionale di Fisica
Nucleare, Sezione di Milano. E-mail: Alberto.Barchielli@polimi.it}}

\vspace*{0.015truein}

\centerline{\footnotesize\it Dipartimento di Matematica, Politecnico di Milano}

\baselineskip=10pt

\centerline{\footnotesize\it Piazza Leonardo da Vinci 32, I-20133 Milano, Italy}

\vspace*{10pt}

\centerline{\footnotesize Giancarlo Lupieri\footnote{Also: Istituto Nazionale di Fisica
Nucleare, Sezione di Milano. E-mail: Giancarlo.Lupieri@mi.infn.it}}

\vspace*{0.015truein} \centerline{\footnotesize\it Dipartimento di Fisica, Universit\`a degli
Studi di Milano}

\baselineskip=10pt

\centerline{\footnotesize\it Via Celoria 16, I-20133 Milano, Italy}

\vspace*{0.225truein}

%\publisher{}{}

\vspace*{0.21truein}

\abstracts{While a positive operator valued measure gives the probabilities in a quantum
measurement, an instrument gives both the probabilities and the a posteriori states. By
interpreting the instrument as a quantum channel and by using the monotonicity theorem for
relative entropies many bounds on the classical information extracted  in a quantum
measurement are obtained in a unified manner. In particular, it is shown that such bounds can
all be stated as inequalities between mutual entropies. This approach based on channels gives
rise to a unified picture of known and new bounds on the classical information (Holevo's,
Shumacher-Westmoreland-Wootters', Hall's, Scutaru's bounds, a new upper bound and a new lower
one). Some examples clarify the mutual relationships among the various bounds.}{}{}

\vspace*{10pt}


\vspace*{3pt}

%\communicate{}

\vspace*{1pt}\textlineskip

\setcounter{footnote}{0}
\renewcommand{\thefootnote}{\alph{footnote}}

\section{\label{intro}Introduction}

A problem which appears in the field of quantum communication and in quantum statistics is the
following: a collection of statistical operators, with some a priori probabilities, describes
the possible states of a quantum system and an observer wants to decide by means of a quantum
measurement in which of these states the system is. The quantity of information extracted by
the measurement is the classical mutual information $I_\rmc$ of the input/output joint
distribution; interesting upper and lower bounds for $I_\rmc$, due to the quantum nature of
the measurement, are given in the literature
\cite{Hol73,JosRW94,Scu95,SchWW96,Hal97,Hal97b,D'A03}.

Usually the measurement is described by a \emph{generalized observable} or \emph{positive
operator valued} (POV) \emph{measure} which allows to obtain the probabilities for the
outcomes of the measurement. However, with respect to a POV measure, a more detailed level of
description of the quantum measurement is represented by a different mathematical object, the
\emph{instrument} \cite{DavL70,Dav76,Oza84}: given a state (the preparation) as input, it
gives as output not only the probabilities of the outcomes but also the state after the
measurement, conditioned on the observed outcome (the a posteriori state). We can think the
instrument to be a channel: from a quantum state (the pre-measurement state) to a
quantum/classical state (a posteriori state plus probabilities). The mathematical
formalization of the idea that an instrument \emph{is} a channel is central in our paper and
allows for a unified approach to various bounds for $I_\rmc$ and for related quantities
\cite{BarL04preA,BarL04preB}.

To maintain things at a sufficiently simple mathematical level, we shall develop and present
all the results in the case of a finite-dimensional Hilbert space, a finite alphabet and an
instrument with finite outcomes.

In Section \ref{sec:instrument} we introduce the notion of instrument and we show how to
associate a channel to it; some inequalities on various relative entropies are deduced from
Ulhmann's monotonicity theorem. From such inequalities we obtain in Section \ref{sec:Hol} some
bounds on the quantity of information $I_\rmc$ which can be extracted by using an instrument
as decoding apparatus; more precisely, we obtain the bound of Holevo \cite{Hol73}
\eqref{Holb}, a slight generalization of the bound of Shumacher, Westmoreland, Wootters (SWW)
\cite{SchWW96} \eqref{SWW} and the new inequalities \eqref{lowbound1}, \eqref{aprioribound}.
From the SWW bound we obtain in a straight way also a result by Groenewold, Lindblad, Ozawa
\cite{Gro71,Lin72,Oza86} on the positivity of the \emph{quantum information gain} given by an
instrument. We also show how such bounds can be stated as inequalities between mutual
entropies (the relative entropy of a bipartite state with respect to its marginals). In
Section \ref{sec:Hall} we generalize a transformation due to Hall \cite{Hal97}, we introduce a
new instrument and we obtain another set of bounds on $I_\rmc$: Hall's bound
\eqref{Halldualbound}, a strengthening of it \eqref{newbound}, Scutaru's bound \cite{Scu95}
\eqref{scu2} and the new inequality \eqref{lastineq}. All the bounds of Sections \ref{sec:Hol}
and \ref{sec:Hall} concern a fixed instrument and the associated POV measure; we can say that
they quantify the performances of the measurement procedure with respect to the initial
ensemble. In Section \ref{sec:example} we give a summary and some examples of the various
bounds.

\section{\label{sec:instrument}Instruments and channels}

Let $\mathcal{H}=\mathbb{C}^d$ be the Hilbert space associated with the quantum system QS; we
denote by $M_d$ the algebra of the complex ($d\! \times\! d$)-matrices and by
$\mathcal{S}_d\subset M_d$ the set of statistical operators on $\mathbb{C}^d$.

\subsection{\label{subsec:finiteInst}Instruments, probabilities and a posteriori states}

We consider a measurement on QS represented by a completely positive instrument $\mathcal{I}$
with finitely many outcomes; let us denote by $\Omega$ the finite set of possible outcomes
(the \emph{value space}). Then, the instrument $\mathcal{I}$ has the structure
\begin{subequations}\label{I+E}
\begin{gather}
\mathcal{I}(F)[\rho]= \sum_{\omega \in F} \mathcal{O}(\omega)[\rho], \qquad \forall F\subset
\Omega, \quad \forall \rho\in M_d,
\\ \label{calO}
\mathcal{O}(\omega)[\rho] = \sum_{k\in K} V_k^\omega \rho V_k^{\omega \dagger} \,,
\\ \label{E&V}
\sum_{\omega \in\Omega} E_{\mathcal{I}}(\omega)=\openone, \qquad
E_{\mathcal{I}}(\omega)=\sum_{k\in K}
 V_k^{\omega \dagger}V_k^\omega,
\end{gather}
\end{subequations}
where  $V_k^\omega\in M_d$, $K$ is a suitable finite set and $\openone$ is the unit element of
$M_d$. Note that $E_{\mathcal{I}}$ is a POV measure, the POV measure associated with
$\mathcal{I}$; $ \mathcal{O}(\omega)$ is an \emph{operation} \cite{Kra83}. If the
pre-measurement state is $\rho\in \mathcal{S}_d$, the probability of the result $\{\omega\in
F\}$, $F\subset \Omega$, is
\begin{equation}\label{probI}
P_\rho(F)=\sum_{\omega \in F} p_\rho(\omega)=  \Tr\{\mathcal{I}(F)[\rho]\}, \qquad
p_\rho(\omega)=\Tr\{E_{\mathcal{I}}(\omega) \rho\} =\Tr\{\mathcal{O}(\omega)[\rho]\},
\end{equation}
and the post-measurement state, conditioned on this result, is
$\big(\Tr\{\mathcal{I}(F)[\rho]\}\big)^{-1}\,\mathcal{I}(F)[\rho]$. When $F$ shrinks to a
single point, the conditional post-measurement state reduces to what is called the \emph{a
posteriori state} \cite{Oza85}
\begin{equation}\label{apostI}
\pi_\rho^{\mathcal{I}}(\omega)=\frac{\mathcal{O}(\omega)[\rho]}{p_\rho(\omega)}, \qquad
\text{if } \ p_\rho(\omega)>0\,;
\end{equation}
this definition has to be completed by defining arbitrarily $\pi_\rho^\mathcal{I}(\omega)$ for
the points $\omega$ for which $p_\rho(\omega)=0$. The a posteriori state is the state to be
attributed to the quantum system QS after the measurement when we know that the result of the
measurement has been exactly $\{\omega\}$. On the opposite side, we have the unconditional
post-measurement state or \emph{a priori state}
\begin{equation}\label{apriori}
\mathcal{I}(\Omega)[\rho]= \sum_{\omega \in \Omega} \mathcal{O}(\omega)[\rho]\,;
\end{equation}
it is the state to be attributed to the system after the measurement, when the result is not
known.

\subsection{\label{subsec:sec}States, entropies, channels}

\subsubsection{Algebras and states}

To formalize the idea that an instrument is a channel, we need to introduce the spaces
$\mathcal{C}(\Omega; M_d)$ of the functions from $\Omega$ into $M_d$ and
$\mathcal{C}(\Omega)\equiv \mathcal{C}(\Omega; \mathbb{C})$, which are finite $C^*$-algebras,
as $M_d$; note that $\mathcal{C}(\Omega; M_d)\simeq \mathcal{C}(\Omega) \otimes M_d$. A state
on a finite $C^*$-algebra is a normalized, positive linear functional on the algebra and in
our cases we have:
\begin{itemize}
\item A state $\rho$ on $M_d$ is identified with a statistical operator,
i.e.\ $\rho\in \mathcal{S}_d$, and $\rho$ applied to an element $a$ of $M_d$ is given by
$\langle \rho, a\rangle = \Tr\{\rho a\}$; this is the usual quantum setup.
\item A state $p$ on $\mathcal{C}(\Omega)$ is a discrete probability density on
$\Omega$ and $\langle p, a\rangle = \sum_{\omega \in \Omega}p(\omega)a(\omega)$; this is the
classical setup.
\item A state $\Sigma$ on $\mathcal{C}(\Omega; M_d)$ is itself an element of
$\mathcal{C}(\Omega; M_d)$ such that $\Sigma(\omega)\geq 0$ and $\sum_{\omega \in
\Omega}\Tr\{\Sigma(\omega)\}=1$; the action of the state $\Sigma$ on an element $a\in
\mathcal{C}(\Omega; M_d)$ is given by $\langle \Sigma, a\rangle =\sum_{\omega \in
\Omega}\Tr\{\Sigma(\omega)a(\omega)\}$. Note the quantum/classical hybrid character of this
case.
\end{itemize}

\subsubsection{Entropies and relative entropies\label{sec:entropies}}

Entropies and relative entropies can be defined in very general situations \cite{OhyP93}, but
here we are interested only in the finite case, where the definitions become simpler. In the
book by Ohya and Petz \cite{OhyP93}, the whole Part I is dedicated to the finite-dimensional
case, while the rest of the book treats the general case. A finite $C^*$-algebra $\mathcal{C}$
can always be seen as a subalgebra of block-diagonal matrices in a big matrix algebra $M_N$
and the definition of entropy for states on $\mathcal{C}$ is derived from the von Neumann
definition for states on $M_N$; the same type of definition applies to the relative entropy
(\cite{OhyP93}, Part I). In some sense this is the general formulation of the trick of
embedding classical probabilities into quantum states, a trick by which many results in
quantum information theory have been proved. Entropies and relative entropies are non
negative; the relative entropy can be infinite. In the case of our three $C^*$-algebras we
have:
\begin{itemize}
\item For $\rho_1,\rho_2 \in \mathcal{S}_d$, the entropy is
\begin{subequations}
\begin{equation}\label{vNentr}
S(\rho_i)= - \Tr\{\rho_i \log \rho_i \}=: S_\rmq(\rho_i)
\end{equation}
(the von Neumann entropy), and  the relative entropy of $\rho_1$ with respect to $\rho_2$ is
\begin{equation}\label{qrelent}
S(\rho_1\|\rho_2)=  \Tr\{\rho_1 (\log \rho_1 - \log \rho_2)\}=:S_\rmq(\rho_1\|\rho_2)\,.
\end{equation}
\end{subequations}
\item In the classical case, for two states $p_1,p_2$ on $\mathcal{C}(\Omega)$, the entropy is
\begin{subequations}
\begin{equation}
S(p_i) = -\sum_{\omega \in \Omega}p_i(\omega)\log p_i(\omega)=:S_\rmc(p_i)
\end{equation}
(the Shannon information), and the relative entropy is
\begin{equation}\label{crelent}
S(p_1\|p_2) = \sum_{\omega \in \Omega}p_1(\omega)\log \frac {p_1(\omega)}{p_2(\omega)}=:
S_\rmc(p_1\|p_2)
\end{equation}
\end{subequations}
(the Kullback-Leibler informational divergence).
\item For two states $\Sigma_1,\Sigma_2$ on $\mathcal{C}(\Omega; M_d)$ we have
\begin{subequations}
\begin{equation}\label{---}
S(\Sigma_i) = - \sum_{\omega \in \Omega}\Tr
\left\{\Sigma_i(\omega)\log\Sigma_i(\omega)\right\} = S_\rmc(p_i)+ \sum_{\omega \in
\Omega}p_i(\omega)S_\rmq\big( \sigma_i(\omega)\big),
\end{equation}
\begin{multline}\label{hybrelent}
S(\Sigma_1\|\Sigma_2) = \sum_{\omega \in \Omega}\Tr \big\{\Sigma_1(\omega)
\big(\log\Sigma_1(\omega)-\log\Sigma_2(\omega)\big)\big\}
\\ {}
= S_\rmc(p_1||p_2)+ \sum_{\omega \in \Omega}p_1(\omega)S_\rmq\big( \sigma_1(\omega) \big\|
\sigma_2(\omega)\big),
\end{multline}
\end{subequations}
\begin{equation}\label{statedecomp}
p_i(\omega):=\Tr \left\{\Sigma_i(\omega)\right\}, \qquad \sigma_i(\omega):= \frac
{\Sigma_i(\omega)}{p_i(\omega)}\,.
\end{equation}
In both equations (\ref{---}) and (\ref{hybrelent}) the first step is by definition and the
second one by simple computations; in \eqref{statedecomp}, when $p_i(\omega)=0$, $
\sigma_i(\omega)$ is defined arbitrarily.
\end{itemize}
In the previous formulas we have used the subscripts ``c'' for ``classical'' and ``q'' for
``quantum'' to underline the cases in which the entropy and the relative entropy are of pure
classical character or of pure quantum one.

\subsubsection{Mutual entropy and $\chi$-quantities.\label{par:me}}

In classical information theory a key concept is that of mutual information which is the
relative entropy of a joint distribution $p_{XY}$ with respect to the product of its marginals
$p_X,\; p_Y$:
\begin{multline}\label{clmutuale}
S_\rmc ( p_{XY} \|p_{X}\otimes p_{Y})  := \sum_{x,y} p_{XY}(x,y)\log
\frac{p_{XY}(x,y)}{p_X(x)p_Y(y)}
\\
{}\equiv \sum_{x} p_X(x)\, S_\rmc(p_{Y|X}(\bullet|x)\|p_Y) \equiv \sum_{y} p_Y(y)\,
S_\rmc(p_{X|Y}(\bullet|y)\|p_X)\,,
\end{multline}
\begin{subequations}
\begin{gather}
p_X(x):= \sum_y p_{XY}(x,y), \qquad p_Y(y):= \sum_x p_{XY}(x,y),
\\
p_{Y|X}(y|x) := \frac{ p_{XY}(x,y)}{p_X(x)}\,, \qquad p_{X|Y}(y|x) := \frac{
p_{XY}(x,y)}{p_Y(y)}\,.
\end{gather}
\end{subequations}

The idea of mutual information can be generalized to all the situations when one has states on
a tensor product of algebras. Let $\mathcal{C}_i$, $i=1,2$ be two finite $C^*$-algebras; let
$\Pi_{12}$ be a state on $\mathcal{C}_1\otimes \mathcal{C}_2$; its \emph{marginals} $\Pi_i$
are its restrictions to the two factors in the tensor product: $\Pi_i :=
\Pi_{12}\big|_{\mathcal{C}_i}$. Then, the \emph{mutual information} or the \emph{mutual
entropy} of the joint state $\Pi_{12}$ is its relative entropy with respect to the tensor
product of its marginals: $S( \Pi_{12}\| \Pi_{1}\otimes \Pi_{2})$.

For instance, in the case $\mathcal{C}_1=\mathcal{C}(\Omega)$, $\mathcal{C}_2= M_d$, a state
$\Sigma$ on $\mathcal{C}_1\otimes \mathcal{C}_2\simeq\mathcal{C}(\Omega;M_d)$ has marginals
$p$ and $\overline \sigma := \sum_\omega \Sigma(\omega) = \sum_\omega p(\omega)
\sigma(\omega)$, where $p(\omega)$ and $\sigma(\omega)$ are defined as in Eq.\
\eqref{statedecomp}. Then, by Eq.\ \eqref{hybrelent} the mutual entropy of $\Sigma$ is
\begin{equation}\label{chi&ent}
S(\Sigma\| p\otimes \overline \sigma) =  \sum_{\omega \in \Omega}p(\omega)S_\rmq\big(
\sigma(\omega) \big\| \overline\sigma\big) \equiv S_\rmq( \overline\sigma)- \sum_{\omega \in
\Omega} p(\omega) S_\rmq\big( \sigma(\omega) \big)
\end{equation}

In quantum information theory, a couple $\{p,\sigma\}$ of a probability $p$ (let us say on the
set $\Omega$) and a family of statistical operators $\sigma(\omega)$ is known as an
\emph{ensemble} and
\begin{equation}
\overline \sigma = \sum_\omega p(\omega) \sigma(\omega)
\end{equation}
is the \emph{average} state of the ensemble. It is trivial to see that the ensemble
$\{p,\sigma\}$ is equivalent to the state  $\Sigma=\{p(\omega)\sigma(\omega)\}$ on
$\mathcal{C}(\Omega;M_d)$; the mutual entropy of this state is called the
\emph{$\chi$-quantity} of the ensemble:
\begin{equation}\label{chi&mutent}
\chi\{p,\sigma\}:= \sum_{\omega \in \Omega}p(\omega)S_\rmq\big( \sigma(\omega) \big\|
\overline\sigma\big)=S(\Sigma\| p\otimes \overline \sigma).
\end{equation}

\subsubsection{Channels}

A (quantum) \emph{channel} $\Lambda$ (\cite{OhyP93} p.\ 137), or dynamical map, or stochastic
map is a completely positive linear map from a finite $C^*$-algebra $\mathcal{C}_1$ to another
one $\mathcal{C}_2$ (but the definition can be extended easily), which transforms states into
states. The composition of channels gives again a channel. Channels are usually introduced to
describe noisy quantum evolutions, but we shall see that also an instrument can be identified
with a channel.

The fundamental \emph{Uhlmann's monotonicity theorem} says that channels decrease the relative
entropy (\cite{OhyP93}, Theor.\ 1.5 p.\ 21): let $\Lambda: \mathcal{C}_1 \to \mathcal{C}_2$ be
a channel between finite $C^*$-algebras; for any two states $\Sigma, \Psi$ on $\mathcal{C}_1$,
the inequality $S(\Sigma\| \Psi)\geq S(\Lambda[\Sigma]\| \Lambda[\Psi])$ holds.

If we have three algebras $\mathcal{A},\mathcal{C}_1,\mathcal{C}_2$ and three channels
$\Lambda_1: \mathcal{A} \to \mathcal{C}_1$, $\Lambda_2:  \mathcal{A} \to \mathcal{C}_2$,
$\Phi: \mathcal{C}_1 \to \mathcal{C}_2$, such that $ \Phi\circ\Lambda_1=\Lambda_2$, we say
that the channel $\Lambda_1$ is a \emph{refinement} of $\Lambda_2$ or that $\Lambda_2$ is a
\emph{coarse graining} of $\Lambda_1$ (\cite{OhyP93} p.\ 138). In this case, for any two
states $\Sigma, \Psi$ on $\mathcal{A}$, we have $S(\Sigma\| \Psi)\geq S(\Lambda_1[\Sigma]\|
\Lambda_1[\Psi])\geq S(\Lambda_2[\Sigma]\| \Lambda_2[\Psi])$.

\subsection{\label{subsec:channel}Instruments, channels and inequalities on relative entropies}

\subsubsection{The instrument as a channel}

Let us define the linear map $\Lambda_{\mathcal{I}}$ from $M_d$ into $\mathcal{C}(\Omega;
M_d)$ by
\begin{equation}\label{channI}
\tau \mapsto \Lambda_{\mathcal{I}}[\tau]\,,  \qquad
\Lambda_{\mathcal{I}}[\tau](\omega):=\mathcal{O}( \omega)[\tau]\,.
\end{equation}
If $\rho \in \mathcal{S}_d$, then $\Lambda_{\mathcal{I}}[\rho]$ is a state on
$\mathcal{C}(\Omega; M_d)$; moreover, by the structure of $\mathcal{O}( \omega)$,
$\Lambda_{\mathcal{I}}$ turns out to be completely positive. Therefore, $
\Lambda_{\mathcal{I}}$ is a channel, the channel associated with the instrument $\mathcal{I}$.
It is also possible to show that any channel from $M_d$ into $\mathcal{C}(\Omega; M_d)$ is the
channel associated to a unique instrument. In the case of general instruments, the
instrument/channel correspondence is treated in \cite{BarL04preB}.

By Uhlmann's monotonicity theorem, we have for any two states $\rho$ and $\phi$ on $M_d$
\begin{equation}\label{Uhll1}
S(\rho\|\phi)\geq S(\Lambda_{\mathcal{I}}[\rho]\|\Lambda_{\mathcal{I}}[\phi])\,.
\end{equation}
By Eqs.\ (\ref{hybrelent}), \eqref{statedecomp}, (\ref{channI}), (\ref{probI}),
(\ref{apostI}), inequality (\ref{Uhll1}) becomes
\begin{equation}\label{!!!}
S_\rmq(\rho\|\phi)\geq S_\rmc(p_\rho\|p_\phi)+ \sum_{\omega\in \Omega}
p_\rho(\omega)S_\rmq\big( \pi_\rho^{\mathcal{I}}(\omega)\big\|
\pi_\phi^{\mathcal{I}}(\omega)\big).
\end{equation}
This is a fundamental inequality. A possible interpretation is that the ``quantum
information'' $S_\rmq(\rho\|\phi)$ contained in the couple of quantum states $\rho$ and $\phi$
is not less than the sum of the classical information $S_\rmc(p_\rho\|p_\phi)$ extracted by
the measurement and of the mean ``quantum information'' $\sum_{\omega\in \Omega}
p_\rho(\omega)S_\rmq\big(\pi_\rho^{\mathcal{I}}(\omega)\big\|
\pi_\phi^{\mathcal{I}}(\omega)\big)$ left in the a posteriori states.

\paragraph*{The POV measure as a channel.}
In \cite{OhyP93}, pp.\ 137-138, another channel is introduced, which involves only the POV
measure, by
\begin{equation}
\Lambda_{E}[\tau](\omega):=\Tr\{E_\mathcal{I} (\omega)\tau\}\,, \qquad \tau \in M_d\,;
\end{equation}
it is easy to check all the properties which define a channel $\Lambda_{E}: M_d \to
\mathcal{C}(\Omega)$. Uhlmann's monotonicity theorem applied to this case gives the inequality
(\cite{OhyP93}, pp.\ 9, 151)
\begin{equation}\label{inOP}
S_\rmq(\rho\|\phi)\geq S_\rmc(p_\rho\|p_\phi)\,,
\end{equation}
which is weaker than (\ref{!!!}). This is due to the fact that inequality (\ref{!!!}) has been
obtained by using a refinement $\Lambda_{\mathcal{I}}$ of the Ohya-Petz channel $\Lambda_{E}$.
Indeed, let us introduce the map $ \Phi_\rmc: \mathcal{C}(\Omega;M_d)\to \mathcal{C}(\Omega)$,
$ \Phi_\rmc[\Sigma](\omega)=\Tr\{ \Sigma(\omega)\}$; in some sense, $\Phi_\rmc$ extracts the
classical part of the state $\Sigma$. Then, it is easy to check that $\Phi_\rmc$ is a channel
and that $\Lambda_{E}= \Phi_\rmc\circ \Lambda_{\mathcal{I}}$.

\subsubsection{The channel $\mathcal{I}(\Omega)$.}

Another inequality is obtained by introducing the channel $\Phi_\rmq$, which extracts the
quantum part of a state $\Sigma$ on $\mathcal{C}(\Omega;M_d)$:
\begin{equation}\label{qchann}
\Phi_\rmq[\Sigma]:= \sum_{\omega\in \Omega} \Sigma(\omega)\,.
\end{equation}
By Eqs.\ (\ref{qchann}), (\ref{channI}), (\ref{apriori}), we get
\begin{equation}
\Phi_\rmq\circ \Lambda_{\mathcal{I}}=\mathcal{I}(\Omega)\,;
\end{equation}
$\mathcal{I}(\Omega)$ is a channel from $M_d$ into itself, which is a coarse graining of
$\Lambda_{\mathcal{I}}$. This gives the inequality
\begin{equation}
S(\Lambda_{\mathcal{I}}[\rho]\|\Lambda_{\mathcal{I}}[\phi]) \geq
S(\mathcal{I}(\Omega)[\rho]\|\mathcal{I}(\Omega)[\phi])
\end{equation}
or
\begin{equation}\label{IOmegabound}
S_\rmc(p_\rho\|p_\phi)+ \sum_{\omega\in \Omega} p_\rho(\omega)S_\rmq\big( \pi_
\rho^\mathcal{I}(\omega)\big\| \pi_\phi^\mathcal{I}(\omega)\big) \geq
S_\rmq(\mathcal{I}(\Omega)[\rho]\|\mathcal{I}(\Omega)[\phi]).
\end{equation}

\subsubsection{\label{para:newbound}A transpose of the channel $\Lambda_E$.}

In \cite{OhyP93} pp.\ 141--143 the transpose of a channel with respect to a fixed state is
defined; such a definition is particularly simple in the case of the channel $\Lambda_E$ and
allows to introduce a new channel which produces new inequalities of interest in quantum
information. Let us fix a quantum state $\phi\in \mathcal{S}_d$, with $p_\phi(\omega)
> 0$, $\forall\omega\in \Omega$; according to \cite{OhyP93}  the $\phi$-transpose of
$\Lambda_E$ is a channel $\Lambda_{E}^\phi : \mathcal{C}(\Omega)\to M_d $, given by
\begin{equation}\label{E_transpose}
\Lambda_{E}^\phi[f]=\sum_{\omega\in\Omega} \frac{f(\omega)}{p_\phi(\omega)} \,
\phi^{1/2}E_{\mathcal{I}}(\omega)\phi^{1/2}\,.
\end{equation}
As it is easy to check, this channel is such that
\begin{equation}
\Lambda_{E}^\phi\circ \Lambda_{E}[\phi]=\Lambda_{E}^\phi[p_\phi]=\phi\,.
\end{equation}
Then, the monotonicity theorem gives
\begin{equation}
S(p_1\|p_2) \geq S(\Lambda_{E}^\phi[p_1]\| \Lambda_{E}^\phi[p_2])\,;
\end{equation}
by taking $p_1=p_\rho$, $p_2=p_\phi$, it becomes
\begin{equation}\label{Etransp}
S_\rmc(p_\rho\|p_\phi) \geq S_\rmq(\Lambda_{E}^\phi[p_\rho]\| \phi).
\end{equation}

\section{\label{sec:Hol}Holevo's bound and related inequalities}

In quantum communication theory often the following scenario is considered: messages are
transmitted by encoding the letters in some quantum states, which are possibly corrupted by a
quantum noisy channel; at the end of the channel the receiver attempts to decode the message
by performing measurements on the quantum system. So, one has an alphabet $A$ (we take it
finite) and the letters $\alpha \in A$ are transmitted with some a priori probabilities
$p_\rmi(\alpha)$; $p_\rmi$ is a discrete probability density on $A$.  Each letter $\alpha$ is
encoded in a quantum state and  we denote by $\rho_\rmi(\alpha)$ the state associated to the
letter $\alpha$ as it arrives to the receiver, after the passage through the transmission
channel. We call these states the letter states and we denote by $\{p_\rmi, \rho_\rmi\}$ the
\emph{ensemble} of the states. We have introduced the subscript ``i'' for ``initial'' and we
shall use ``f'' for final.

Let us use the instrument $\mathcal{I}$, given in Section \ref{subsec:finiteInst}, as decoding
apparatus. The conditional probability of the outcome $\omega$, given the input letter
$\alpha$, is
\begin{subequations}
\begin{equation}
p_{\rmf|\rmi}(\omega|\alpha)= \Tr\{ \mathcal{O}(\omega)[\rho_\rmi(\alpha)]\} \equiv \Tr\{
E_{\mathcal{I}}(\omega)\rho_\rmi(\alpha)\}\,;
\end{equation}
then, the joint probability of input and output, the conditional probability of the input
given the output and the marginal probability of the output are given by
\begin{gather}
p_{\rmi\rmf}( \alpha, \omega)=p_{\rmf|\rmi}(\omega|\alpha)p_{\rmi}(\alpha)\,, \qquad
p_{\rmi|\rmf}(\alpha|\omega) = \frac{p_{\rmf|\rmi}(
\omega|\alpha)p_\rmi(\alpha)}{p_\rmf(\omega)}\,,
\\
p_{\rmf}(\omega)=\sum_\alpha p_{\rmi\rmf}( \alpha, \omega) =\sum_\alpha
p_{\rmi}(\alpha)\,\Tr\{ \mathcal{O}( \omega)[\rho_\rmi(\alpha)]\}=\Tr\{ \mathcal{O}(
\omega)[\eta_\rmi]\}\,,
\end{gather}
\end{subequations}
where $\eta_\rmi$ is the average state of the initial ensemble, or \emph{initial a priori
state}:
\begin{equation}\label{iapriori}
\eta_\rmi:= \sum_{\alpha\in A} p_\rmi(\alpha)\,  \rho_\rmi(\alpha).
\end{equation}
Note that $p_{\rmi|\rmf}(\alpha|\omega)$ is well defined only when $p_\rmf(\omega)>0$, but it
can be arbitrarily completed  when $p_\rmf(\omega)=0$.

The mean information $I_\rmc\{p_\rmi, \rho_\rmi;E_{\mathcal{I}}\}$ on the transmitted letter
which can be extracted in this way is the input/output classical mutual information, cf.\
\eqref{clmutuale}:
\begin{equation}
I_\rmc\{p_\rmi, \rho_\rmi;E_{\mathcal{I}}\} := S_\rmc ( p_{\rmi\rmf} \|p_{\rmi}\otimes
p_{\rmf}) = \sum_{\alpha} p_\rmi(\alpha)\, S_\rmc(p_{\rmf|\rmi}(\bullet|\alpha)\|p_\rmf)\,.
\end{equation}

\subsection{Holevo's upper bound and the ``transpose channel'' lower bound}

\subsubsection{Holevo's bound}

Let us introduce \emph{Holevo's $\chi$-quantity}, i.e.\ the $\chi$-quantity of the initial
ensemble (cf.\ Eqs.\ \eqref{chi&ent}--\eqref{chi&mutent})
\begin{equation}\label{chi}
\chi\{p_{\rmi},\rho_\rmi\}:= \sum_{\alpha\in A} p_{\rmi}(\alpha)\,
S_\rmq(\rho_\rmi(\alpha)\|\eta_\rmi)\\ {}= S_\rmq(\eta_\rmi)- \sum_{\alpha\in A}
p_{\rmi}(\alpha) S_\rmq(\rho_\rmi(\alpha))\,.
\end{equation}
By applying the inequality (\ref{inOP}) to the states $\rho_\rmi(\alpha)$ and $\eta_\rmi$ and
then by multiplying by $p_\rmi(\alpha)$ and summing on $\alpha$, one gets Holevo's inequality
\cite{Hol73}
\begin{equation}\label{Holb}
I_\rmc \{p_\rmi, \rho_\rmi;E_{\mathcal{I}}\} \leq \chi\{p_{\rmi},\rho_\rmi\}\,.
\end{equation}
In the case of a general Hilbert space, general POV measure, general alphabet, this inequality
has been proved, just by using the channel $\Lambda_E$, by Yuen and Ozawa in \cite{YueO93}.

\subsubsection{The lower bound}

The monotonicity theorem applied to the channel $\Lambda_E^{\eta_\rmi}$, the
$\eta_\rmi$-transpose of $\Lambda_E$, gives a new lower bound for $I_\rmc$.

Firstly, from \eqref{E_transpose} one has
\begin{equation}
\Lambda_E^{\eta_\rmi}[f]= \sum_\omega f(\omega) \sigma(\omega),
\end{equation}
where we have introduced the family of statistical operators
\begin{equation}\label{newinistat}
\sigma(\omega):= \frac 1 {p_\rmf(\omega)}\, \eta_\rmi^{1/2}
E_{\mathcal{I}}(\omega)\eta_\rmi^{1/2}\,.
\end{equation}
The probability $p_\rmf(\omega)$ could vanish for some $\omega$'s, but in this case the
positivity implies that also $\eta_\rmi^{1/2} E_{\mathcal{I}}(\omega)\eta_\rmi^{1/2}$ vanishes
and the definition above can be completed arbitrarily for such $\omega$'s. Note that the
ensemble $\{p_\rmf,\sigma\}$ has average
\begin{equation}
\sum_\omega p_\rmf(\omega)\sigma(\omega)=\eta_\rmi\,.
\end{equation}

Then, by applying the inequality (\ref{Etransp}) to the states $\rho_\rmi(\alpha)$ and
$\eta_\rmi$, by multiplying by $p_\rmi(\alpha)$ and summing on $\alpha$, one gets
\begin{equation}\label{lowbound1}
I_\rmc \{p_\rmi, \rho_\rmi;E_{\mathcal{I}}\} \geq \chi\{p_\rmi ,\xi\},
\end{equation}
where
\begin{equation}\label{statesxi}
\xi(\alpha):= \sum_\omega p_{\rmf|\rmi}(\omega|\alpha)\sigma(\omega)\,.
\end{equation}
The ensemble $\{p_\rmi,\xi\}$ has average
\begin{equation}
\sum_\alpha p_\rmi(\alpha)\xi(\alpha)=\eta_\rmi\,.
\end{equation}

It is possible to show that, according to the definition of transpose given in Ref.\
\cite{OhyP93}, the $p_\rmf$-transpose of $\Lambda_E^{\eta_\rmi}$ would be $\Lambda_E$.
Therefore, there is a sort of duality between the channels $\Lambda_E$ and
$\Lambda_E^{\eta_\rmi}$ and, so, between Holevo's bound \eqref{Holb} and the bound
\eqref{lowbound1}.

\subsection{The bound of Schumacher, Westmoreland, Wootters}

Let us consider now the a posteriori states
\begin{equation}
\rho^\alpha_\rmf(\omega):= \pi_{\rho_\rmi(\alpha)}^{\mathcal{I}}(\omega)=
\frac{\mathcal{O}(\omega)[\rho_\rmi(\alpha)]}{p_{\rmf|\rmi}(\omega|\alpha)} \,, \qquad
\rho_\rmf(\omega) := \pi_{\eta_\rmi}^{\mathcal{I}}(\omega)=
\frac{\mathcal{O}(\omega)[\eta_\rmi]}{p_{\rmf}(\omega)}\,.
\end{equation}
By applying the inequality (\ref{!!!}) to the states $\rho_\rmi(\alpha)$ and $\eta_\rmi$ and
then by multiplying by $p_\rmi(\alpha)$ and summing on $\alpha$, one gets
\begin{equation}\label{SWW}
\chi\{p_{\rmi},\rho_\rmi\} \geq I_\rmc \{p_\rmi, \rho_\rmi;E_{\mathcal{I}}\}  + \sum_{\omega}
p_{\rmf}(\omega) \, \chi\{p_{\rmi|\rmf}(\bullet|\omega),\rho_\rmf^\bullet(\omega)\}.
\end{equation}
The average state of the ensemble
$\{p_{\rmi|\rmf}(\bullet|\omega),\rho_\rmf^\bullet(\omega)\}$ is
\begin{equation}
\sum_\alpha p_{\rmi|\rmf}( \alpha|\omega) \,\rho^\alpha_\rmf(\omega)= \rho_{\rmf} (\omega).
\end{equation}

Note that
\begin{equation}
\sum_{\omega} p_{\rmf}(\omega) \,
\chi\{p_{\rmi|\rmf}(\bullet|\omega),\rho_\rmf^\bullet(\omega)\}\equiv \sum_{\omega}
p_{\rmf}(\omega) \,S_\rmq\big(\rho_\rmf(\omega)\big)-\sum_{\alpha,\omega}
p_{\rmi\rmf}(\alpha,\omega)\,S_\rmq\big(\rho_\rmf^\alpha(\omega)\big)
\end{equation}
is the mean $\chi$-quantity left in the a posteriori states by the instrument. Inequality
(\ref{SWW}) gives an upper bound on $I_\rmc\{p_\rmi, \rho_\rmi;E_{\mathcal{I}}\} $ stronger
than (\ref{Holb}); indeed, the extra term vanishes when $\rho^\alpha_\rmf(\omega)$ is almost
surely independent from $\alpha$, as in the case of a von Neumann complete measurement, but
for a generic instrument it is positive.

The original SWW bound \cite{SchWW96} is inequality (\ref{SWW}) in the case of an instrument
with no sum on $k$ in the definition \eqref{calO} of the operations $\mathcal{O}(\omega)$.
Eq.\ (\ref{SWW}) is a slight generalization to the case of \eqref{calO} with sums and was
already proven in \cite{BarL04preA}; a different proof, more similar to the SWW original one,
was given after in \cite{Jac04}. Inequality (\ref{SWW}) has been generalized to the infinite
and continuous case in \cite{BarL04preB}.

Roughly, Eq.\ (\ref{SWW}) says that the quantum information contained in the initial ensemble
$\{p_{\rmi},\rho_\rmi\}$ is greater than the classical information extracted in the
measurement plus the mean quantum information left in the a posteriori states. Inequality
(\ref{SWW}) can be seen also as giving some kind of information/disturbance trade-off, a
subject to which the paper \cite{D'A03}, which contains a somewhat related inequality, is
devoted.

Let us introduce the \emph{a priori final states}
\begin{subequations}\label{apriorifinal}
\begin{gather}\label{apriorifinal1}
\eta^\alpha_\rmf
:=\mathcal{I}(\Omega)[\rho_\rmi(\alpha)]=\sum_{\omega}\mathcal{O}(\omega)[\rho_\rmi(\alpha)]
=\sum_{\omega} p_{\rmf|\rmi}(\omega|\alpha)\, \rho^\alpha_\rmf(\omega),
\\ \label{apriorifinal2}
\eta_\rmf :=\mathcal{I}(\Omega)[\eta_\rmi] =\sum_{\omega}\mathcal{O}(\omega)[\eta_\rmi] =
\sum_{\omega} p_{\rmf}(\omega)\, \rho_\rmf(\omega) =\sum_{\alpha, \omega} p_{\rmi\rmf}(
\alpha, \omega) \, \rho^\alpha_\rmf(\omega) = \sum_{\alpha }p_\rmi(\alpha) \,
\eta^\alpha_\rmf\,.
\end{gather}
\end{subequations}
By using the expression of a $\chi$-quantity in terms of entropies
\eqref{chi&ent}--\eqref{chi&mutent}, one can check that the following identity holds
\begin{equation}\label{idt1}
\chi\{p_{\rmi\rmf},\rho_{\rmf}^\bullet\}= \chi\{p_{\rmf},\rho_{\rmf}\}+ \sum_\omega p_\rmf(
\omega)\, \chi\big\{p_{\rmi|\rmf}(\bullet|\omega),\rho_{\rmf}^\bullet (\omega)\big\}.
\end{equation}
Both the new ensembles $\{p_{\rmi\rmf},\rho_{\rmf}^\bullet\}$ and $\{p_{\rmf},\rho_{\rmf}\}$
have $\eta_\rmf$ as average state. By using this identity, inequality (\ref{SWW}) can be
rewritten in the slightly more symmetric equivalent form
\begin{equation}\label{BL1}
I_\rmc\{p_\rmi,\rho_\rmi;E_\mathcal{I}\} \leq \chi\{p_{\rmi},\rho_\rmi\} +
\chi\{p_{\rmf},\rho_{\rmf}\} - \chi\{p_{\rmi\rmf},\rho_{\rmf}^\bullet\}.
\end{equation}

\subsection{The generalized Groenewold-Lindblad inequality}

Given an instrument $\mathcal{I}$ and a statistical operator $\eta$, an interesting quantity,
which can be called the \emph{quantum information gain}, is
\begin{equation}
I_\rmq(\eta;\mathcal{I})= S_\rmq(\eta) - \sum_\omega
S_\rmq\big(\pi_\eta^{\mathcal{I}}(\omega)\big)\, p_{\eta}( \omega)\,;
\end{equation}
this is nothing but the entropy of the pre-measurement state minus the mean entropy of the a
posteriori states.

By using the expression of a $\chi$-quantity in terms of entropies and mean entropies, as in
Eq.\ (\ref{chi}), one can see that inequality (\ref{SWW}) is equivalent to
\begin{equation}\label{Iqineq}
I_\rmq(\eta_\rmi;\mathcal{I}) -\sum_\alpha p_{\rmi}(\alpha)\,
I_\rmq(\rho_\rmi(\alpha);\mathcal{I})\geq I_\rmc\{p_\rmi, \rho_\rmi;E_{\mathcal{I}}\}  \geq0
\,.
\end{equation}
Note that, once the instrument is fixed, $I_\rmq(\eta_\rmi;\mathcal{I})$ depends only on
$\eta_\rmi$, while both $I_\rmc\{p_\rmi, \rho_\rmi;E_{\mathcal{I}}\} $ and $\sum_\alpha
p_{\rmi}(\alpha)\,I_\rmq(\rho_\rmi(\alpha);\mathcal{I})$ depend on the demixture
$\{p_\rmi,\rho_\rmi\}$ of $\eta_\rmi$.

An interesting question is when the quantum information gain is positive. Groenewold has
conjectured \cite{Gro71} and Lindblad \cite{Lin72} has proved that the quantum information
gain is non negative for an instrument of the von Neumann-L\"uders type. The general case has
been settled down by Ozawa, who has introduced the a posteriori states for general instruments
in \cite{Oza85} and in \cite{Oza86} has proved a general result on instruments preserving pure
states, which here we state only in the finite dimensional and discrete case.

\begin{theorem} \label{TheoOza}
For an instrument $\mathcal{I}$ as in Eq.\ \emph{(\ref{I+E})}, the two following statements
are equivalent:
\begin{description}
\item[\textrm{(a)}] the instrument $\mathcal{I}$ sends any pure input state into almost surely pure a posteriori states;
\item[\textrm{(b)}] $I_\rmq(\eta;\mathcal{I}) \geq 0$, for all statistical operators
$\eta$.
\end{description}
\end{theorem}
Now the proof is an easy application of inequality (\ref{Iqineq}); this proof works also in
the general case \cite{BarL04preB}.

\smallskip

\noindent\textit{Proof}. To prove that (b) implies (a) is trivial; it is enough to put a pure
state $\eta$ into the definition, which gives
\[
0\leq I_\rmq(\eta;\mathcal{I})=  - \sum_\omega
S_\rmq\big(\pi_\eta^{\mathcal{I}}(\omega)\big)\, p_{\eta}( \omega)\,.
\]
This implies that the a posteriori states $\pi_\eta^{\mathcal{I}}(\omega)$ are $p_\eta$-almost
surely pure, because the von Neumann entropy vanishes only on the pure states.

To show that (a) implies (b), the non trivial part in Ozawa's proof, let $\eta_\rmi$ be a
generic state and $\{p_\rmi,\rho_\rmi\}$ be a demixture of it into pure states; then, by (a)
$I_\rmq(\rho_\rmi(\alpha);\mathcal{I})=0$ and (\ref{Iqineq}) reduces to
$I_\rmq(\eta_\rmi;\mathcal{I}) \geq I_\rmc\{p_\rmi, \rho_\rmi;E_{\mathcal{I}}\} \geq 0$, which
is (b). \hfill{$\square$}

\medskip

A sufficient condition for $\mathcal{I}$ being a pure state preserving instrument is to take
$|K|=1$ in \eqref{calO}, but this is not necessary. The complete characterization of the
structure of a pure state preserving instrument has been given in \cite{Oza95}.

Inequality (\ref{Iqineq}) is also interesting in itself, because it gives a link between the
quantum information gain in the case of a pre-measurement state $\eta_\rmi$ and the mean
quantum information gain in the case of a demixture of $\eta_\rmi$, a link which holds true
for any kind of instrument. The amount of quantum information has been studied and its meaning
discussed also in \cite{Jac03,Jac04}, where also the connections with inequality (\ref{SWW})
and with pure state preserving instruments have been pointed out.

\subsection{Post-measurement $\chi$-quantities}

By applying the inequality (\ref{IOmegabound}) to the states $\rho_\rmi(\alpha)$ and
$\eta_\rmi$ and then by multiplying by $p_\rmi(\alpha)$ and summing on $\alpha$, one gets
\begin{equation}\label{aprioribound}
I_\rmc\{p_\rmi, \rho_\rmi;E_{\mathcal{I}}\}  +\sum_{\omega} p_{\rmf}(\omega) \,
\chi\{p_{\rmi|\rmf}(\bullet|\omega),\rho^\bullet_\rmf(\omega)\} \geq
\chi\{p_{\rmi},\eta_\rmf^\bullet\}.
\end{equation}
By Eqs.\ \eqref{apriorifinal} the average state of the ensemble
$\{p_{\rmi},\eta_\rmf^\bullet\}$ is $\eta_\rmf$.

Similarly to \eqref{idt1}, also a second identity holds:
\begin{equation}\label{idt2}
\chi\{p_{\rmf},\rho_{\rmf}\}+ \sum_\omega p_\rmf( \omega)\,
\chi\big\{p_{\rmi|\rmf}(\bullet|\omega),\rho_{\rmf}^\bullet (\omega)\big\}
\\
{}= \chi\{p_{\rmi},\eta_{\rmf}^\bullet\}+ \sum_\alpha p_\rmi(\alpha)\,
\chi\big\{p_{\rmf|\rmi}(\bullet|\alpha),\rho_{\rmf}^\alpha\big\}.
\end{equation}
By \eqref{apriorifinal1}, the ensemble
$\big\{p_{\rmf|\rmi}(\bullet|\alpha),\rho_{\rmf}^\alpha\big\}$ has average state
$\eta_\rmf^\alpha$. By this identity, inequality \eqref{aprioribound} is equivalent to
\begin{equation}\label{varineq}
I_\rmc\{p_\rmi,\rho_\rmi;E_\mathcal{I}\} + \sum_\alpha p_\rmi(\alpha)\,
\chi\big\{p_{\rmf|\rmi}(\bullet|\alpha),\rho_{\rmf}^\alpha\big\}
\\ {}\geq
\chi\{p_{\rmf},\rho_{\rmf}\}.
\end{equation}

\subsection{Mutual entropy formulation\label{subsec:mutent}}

\subsubsection{The initial and the final state}

Let us introduce the algebras
\begin{equation}
\mathcal{C}_0 := \mathcal{C}(A), \qquad \mathcal{C}_1:= M_d, \qquad \mathcal{C}_2:=
\mathcal{C}(\Omega).
\end{equation}
As seen in Paragraph \ref{par:me}, the initial ensemble $\{p_\rmi,\rho_\rmi\}$ can be seen as
a state $\Sigma_\rmi^{01}$ on $\mathcal{C}_0\otimes \mathcal{C}_1\simeq \mathcal{C}(A;M_d)$.
By using a superscript which indicates the algebras on which a state is acting, we can write
\begin{equation}\label{initialstate}
\Sigma_\rmi^{01}:=\{p_\rmi(\alpha)\rho_\rmi(\alpha)\}, \qquad
\Sigma_\rmi^0=\{p_\rmi(\alpha)\}, \qquad \Sigma_\rmi^1=\{\eta_\rmi\},
\end{equation}
for the initial state and its marginals. By \eqref{chi&mutent}, Holevo's $\chi$-quantity
\eqref{chi} coincides with the initial mutual entropy
\begin{equation}\label{initmentr}
S(\Sigma_\rmi^{01}\|\Sigma_\rmi^0\otimes \Sigma_\rmi^1)= \chi\{p_{\rmi},\rho_\rmi\}.
\end{equation}

By dilating the channel $\Lambda_\mathcal{I}$ \eqref{channI}  with the identity we obtain the
\emph{measurement channel}
\begin{equation}
\Lambda : \mathcal{C}_0\otimes \mathcal{C}_1 \to \mathcal{C}_0\otimes \mathcal{C}_1 \otimes
\mathcal{C}_2 \,, \qquad \Lambda := \openone \otimes \Lambda_{\mathcal{I}}\,.
\end{equation}
Then, by applying the measurement channel to the initial state we obtain the final state
\begin{subequations}
\begin{equation}\label{defsigmaf}
\Sigma^{012}_\rmf:= \Lambda[\Sigma^{01}_\rmi] \\ {}= \{p_\rmi(\alpha)
\Lambda_{\mathcal{I}}[\rho_\rmi(\alpha)](\omega)\} = \{p_{\rmi\rmf}(\alpha,\omega)
\rho_{\rmf}^{\alpha}(\omega)\},
\end{equation}
whose marginals are
\begin{equation}
\begin{aligned}
& \Sigma^{01}_\rmf= \{p_\rmi(\alpha)\eta_{\rmf}^{\alpha}\}, &\qquad & \Sigma^{02}_\rmf=
\{p_{\rmi\rmf}(\alpha,\omega) \}, &\qquad & \Sigma^{12}_\rmf= \{
p_\rmf(\omega)\rho_{\rmf}(\omega)\}, \\ & \Sigma^{0}_\rmf= \Sigma^{0}_\rmi=\{p_\rmi(\alpha)\},
&\qquad & \Sigma^{1}_\rmf= \{\eta_\rmf \}, &\qquad & \Sigma^{2}_\rmf= \{ p_\rmf(\omega)\}.
\end{aligned}
\end{equation}
\end{subequations}
Moreover, one gets easily
\begin{equation}\label{tensf}
\Lambda[\Sigma^{0}_\rmi\otimes \Sigma^{1}_\rmi]=\Sigma^{0}_\rmf\otimes \Sigma^{12}_\rmf\,.
\end{equation}

\subsubsection{Mutual entropies and inequalities}

By the definitions of Section \ref{sec:entropies} it is easy to compute all the mutual
entropies related to the final state. The mutual entropy involving only the classical part of
the final state turns out to be the input/output classical mutual information:
\begin{equation}\label{fme02}
S(\Sigma_\rmf^{02}\|\Sigma_\rmf^0\otimes \Sigma_\rmf^{2})= S_\rmc(p_{\rmi\rmf}\|p_\rmi\otimes
p_\rmf)= I_\rmc\{p_\rmi,\rho_\rmi;E_\mathcal{I}\}.
\end{equation}
Then, the remaining mutual entropies turn out to be
\begin{subequations}
\begin{equation}\label{fme01}
S(\Sigma_\rmf^{01}\|\Sigma_\rmf^0\otimes \Sigma_\rmf^1)= \chi\{p_{\rmi},\eta_{\rmf}^\bullet\},
\qquad S(\Sigma_\rmf^{12}\|\Sigma_\rmf^1\otimes \Sigma_\rmf^2)= \chi\{p_{\rmf},\rho_{\rmf}\},
\end{equation}
\begin{equation}\label{fme02,1}
S(\Sigma_\rmf^{012}\|\Sigma_\rmf^{02}\otimes \Sigma_\rmf^{1})=
\chi\{p_{\rmi\rmf},\rho_{\rmf}^\bullet\},
\end{equation}
\begin{equation}\label{fme0,12}
S(\Sigma_\rmf^{012}\|\Sigma_\rmf^0\otimes \Sigma_\rmf^{12})=
I_\rmc\{p_\rmi,\rho_\rmi;E_\mathcal{I}\} \\ {}+ \sum_\omega p_\rmf(\omega)\,
\chi\big\{p_{\rmi|\rmf}(\bullet|\omega),\rho_{\rmf}^\bullet (\omega)\big\},
\end{equation}
\begin{equation}\label{fme01,2}
S(\Sigma_\rmf^{012}\|\Sigma_\rmf^{01}\otimes \Sigma_\rmf^{2})=
I_\rmc\{p_\rmi,\rho_\rmi;E_\mathcal{I}\} \\ {}+ \sum_\alpha p_\rmi(\alpha)\,
\chi\big\{p_{\rmf|\rmi}(\bullet|\alpha),\rho_{\rmf}^\alpha\big\},
\end{equation}
\begin{equation}\label{fme0,1,2}
S(\Sigma_\rmf^{012}\|\Sigma_\rmf^{0}\otimes \Sigma_\rmf^{2}\otimes \Sigma_\rmf^{3}) =
I_\rmc\{p_\rmi,\rho_\rmi;E_\mathcal{I}\} + \chi\{p_{\rmi\rmf},\rho_{\rmf}^\bullet\}.
\end{equation}
\end{subequations}
Note that the expressions of the mutual entropies involve the $\chi$-quantities of all the
ensembles entering into play.

Uhlmann's monotonicity theorem and Eqs.\ (\ref{defsigmaf}), (\ref{tensf}) give us the
inequality
\begin{equation}
S(\Sigma_\rmi^{01}\|\Sigma_\rmi^0\otimes \Sigma_\rmi^1)\geq
S(\Lambda[\Sigma_\rmi^{01}]\|\Lambda[\Sigma_\rmi^0\otimes \Sigma_\rmi^1]) =
S(\Sigma_\rmf^{012}\|\Sigma_\rmf^0\otimes \Sigma_\rmf^{12}).
\end{equation}
By Eqs.\ \eqref{initmentr} and \eqref{fme0,12}, one has that this inequality is equivalent to
the SWW bound \eqref{SWW}.

It is trivial to see that the operation of restricting states on a tensor product to one of
the factors is a channel; therefore, we have also the inequality
\begin{equation}
S(\Sigma_\rmf^{012}\|\Sigma_\rmf^0\otimes \Sigma_\rmf^{12})\geq
S(\Sigma_\rmf^{01}\|\Sigma_\rmf^0\otimes \Sigma_\rmf^{1}),
\end{equation}
which, by \eqref{fme0,12} and \eqref{fme01}, is equivalent to inequality \eqref{aprioribound}.
All the other inequalities which can be obtained are implied by the previous ones trivially or
via the identities (\ref{idt1}), \eqref{idt2}. Among these inequalities there is
\begin{equation}
S(\Sigma_\rmi^{01}\|\Sigma_\rmi^0\otimes \Sigma_\rmi^1)\geq
S(\Sigma_\rmf^{02}\|\Sigma_\rmf^0\otimes \Sigma_\rmf^{2}),
\end{equation}
which, by \eqref{chi&mutent}, \eqref{fme02}, is equivalent to Holevo's bound \eqref{Holb}.

To express inequality \eqref{lowbound1} in terms of mutual entropies let us introduce the new
channel
\begin{subequations}
\begin{equation}
\Gamma : \mathcal{C}_0\otimes\mathcal{C}_2 \to \mathcal{C}_0\otimes\mathcal{C}_1
\end{equation}
by
\begin{equation}
\Gamma[f](\alpha) = \sum_\omega f(\alpha,\omega) \sigma(\omega), \qquad\forall f \in
\mathcal{C}_0\otimes\mathcal{C}_2\,.
\end{equation}
\end{subequations}
Then, the monotonicity theorem gives
\begin{equation}\label{nmei}
S(p_{\rmi\rmf}\|p_\rmi\otimes p_\rmf) \geq S(\Gamma[p_{\rmi\rmf}]\|\Gamma[p_\rmi\otimes
p_\rmf]);
\end{equation}
but one has
\begin{subequations}
\begin{gather}
\Gamma[p_{\rmi\rmf}](\alpha)= \sum_\omega p_{\rmi\rmf}(\alpha,\omega)
\sigma(\omega)=p_\rmi(\alpha) \xi(\alpha)\,,
\\
\Gamma[p_{\rmi}\otimes p_{\rmf}](\alpha)= \sum_\omega p_{\rmi}(\alpha) p_{\rmf}(\omega)
\sigma(\omega)=p_\rmi(\alpha) \eta_\rmi\,,
\end{gather}
\end{subequations}
and, so, inequality \eqref{nmei} is equivalent to the bound \eqref{lowbound1}. Note that
$\Gamma[p_{\rmi}\otimes p_{\rmf}]= \Gamma[p_{\rmi\rmf}]\big|_{\mathcal{C}_0} \otimes
\Gamma[p_{\rmi\rmf}]\big|_{\mathcal{C}_1}$ so that both sides of \eqref{nmei} are mutual
entropies.

\section{Hall's bound and generalizations \label{sec:Hall}}

In \cite{Hal97} Hall exhibits a transformation on the initial ensemble and on the POV measure
which leaves invariant $I_\rmc$ but not the initial $\chi$-quantity and in this way produces a
new upper bound on the classical information. Inspired by Hall's transformation, a new
instrument can be constructed in such a way that the analogous of inequality (\ref{SWW})
produces an upper bound on $I_\rmc$ stronger than both Hall's and Holevo's ones.

For simplicity in this section we assume that $\eta_\rmi$ is invertible.

\subsection{A generalization of Hall's transformation}

\subsubsection{A new instrument $\mathcal{J}$}

Let us set
\begin{subequations}
\begin{equation}\label{Malpha}
M(\alpha):= \sqrt{p_\rmi(\alpha)}\ \rho_\rmi(\alpha)^{1/2}\eta_\rmi^{-1/2}\,, \qquad
\mathcal{G}(\alpha)[\tau] := M(\alpha) \tau M(\alpha)^*\,, \quad \forall \tau \in M_d\,;
\end{equation}
by Eq.\ (\ref{iapriori}) the operators $M(\alpha)$ satisfy the normalization condition
\begin{equation}
\sum_\alpha M(\alpha)^*M(\alpha)=\openone\,.
\end{equation}
Then, the position
\begin{equation}
\mathcal{J}(B) := \sum_{\alpha\in B} \mathcal{G}(\alpha),\qquad B\subset A,
\end{equation}
defines an instrument with value space $A$. The instrument $\mathcal{J}$ has been constructed
by using only the old initial ensemble $\{p_\rmi, \rho_\rmi\}$. The associated POV measure is
\begin{equation}\label{newE}
E_\mathcal{J}(\alpha) = M(\alpha)^*M(\alpha)= p_\rmi(\alpha)\, \eta_\rmi^{-1/2}
\rho_\rmi(\alpha)\eta_\rmi^{-1/2}\,.
\end{equation}
\end{subequations}

Now, we can construct the associated channel and a posteriori states, as in Section
\ref{sec:instrument}: $\forall \tau \in M_d$, $\forall \rho \in \mathcal{S}_d$, one has
\begin{equation}
\Lambda_\mathcal{J}[\tau](\alpha)=\mathcal{G}(\alpha)[\tau]=M(\alpha) \tau M(\alpha)^*,
\end{equation}
\begin{equation}\label{apostalpha}
\pi_\rho^{\mathcal{J}}(\alpha)= \big(\Tr \left\{M(\alpha)^*M(\alpha)\rho \right\} \big)^{-1}
M(\alpha)\rho M(\alpha)^* .
\end{equation}
Let us stress that $\mathcal{J}$ sends pure states into a.s.\ pure a posteriori states;
therefore, by Theorem \ref{TheoOza} one has
\begin{equation}\label{newIq}
I_\rmq\{\rho;\mathcal{J}\} \equiv S_\rmq(\rho) - \sum_\alpha
\Tr\{E_\mathcal{J}(\alpha)\rho\}\, S_\rmq \big( \pi_\rho^\mathcal{J}(\alpha)\big)\geq 0.
\end{equation}

\subsubsection{A new initial ensemble and the replacements}

Now we consider $\{p_\rmf,\sigma\}$ \eqref{newinistat} as initial ensemble for $\mathcal{J}$;
recall that its average state is $\eta_\rmi$ \eqref{iapriori}. It is easy to verify that
\begin{equation}
\Tr\{E_\mathcal{J}(\alpha)\sigma(\omega)\}= p_{\rmi|\rmf}( \alpha|\omega);
\end{equation}
together with the substitution of $p_\rmi$ with $p_\rmf$, this gives that $p_{\rmi\rmf}$ is
left invariant and that $p_\rmf$ is substituted by $p_\rmi$. Therefore, we have
\begin{equation}\label{Iinvariance}
I_\rmc\{p_\rmf,\sigma;E_\mathcal{J}\} =I_\rmc\{p_\rmi,\rho_\rmi;E_\mathcal{I}\}.
\end{equation}
Indeed, the POV measure $E_\mathcal{J}$ and the states $\sigma(\omega)$ have been constructed
by Hall just in order to have this equality.

One can also check that under Hall's transformation the states $\sigma(\omega)$
\eqref{newinistat} become the states $\rho_\rmi(\alpha)$. Summarizing, we have that the
following replacements have to be made:
\begin{subequations}\label{79's}
\begin{equation}\label{rho2sigma}
\begin{split}
A \rightleftarrows \Omega\,, \qquad &p_{\rmi\rmf} \rightarrow p_{\rmi\rmf}\,, \qquad
p_\rmi(\alpha) \rightleftarrows p_\rmf(\omega)\,, \\ p_{\rmf|\rmi}(\omega|\alpha)
\rightleftarrows p_{\rmi|\rmf}(\alpha|\omega)\,, \qquad &\rho_\rmi(\alpha) \rightleftarrows
\sigma(\omega)\,, \qquad  \eta_\rmi \rightarrow \eta_\rmi\,.
\end{split}
\end{equation}
By Eqs.\ (\ref{newinistat}), (\ref{Malpha}), (\ref{apostalpha}) we obtain also
\begin{equation} \label{apostJ1}
\rho_\rmf^\alpha(\omega) \rightarrow \pi^\mathcal{J}_{\sigma(\omega)}(\alpha)=
\rho_\rmi(\alpha)^{1/2} \, \frac{E_\mathcal{I}(\omega)
}{p_{\rmf|\rmi}(\omega|\alpha)}\,\rho_\rmi(\alpha)^{1/2}\,, \qquad \rho_\rmf(\omega)
\rightarrow\pi_{\eta_\rmi}^\mathcal{J}(\alpha)= \rho_\rmi(\alpha);
\end{equation}
the first quantity is defined similarly to (\ref{newinistat}). Moreover,
\begin{equation}\label{newapr}
\eta_\rmf^\alpha \rightarrow \eta^\omega_\mathcal{J} :=\sum_\alpha
p_{\rmi|\rmf}(\alpha|\omega) \pi^\mathcal{J}_{\sigma(\omega)}(\alpha) = \sum_\alpha
\frac{p_\rmi(\alpha) }{p_{\rmf}(\omega)}\,\rho_\rmi(\alpha)^{1/2} E_\mathcal{I}(\omega)
\rho_\rmi(\alpha)^{1/2}\,,
\end{equation}
\begin{equation}
\eta_\rmf \rightarrow \sum_\alpha p_\rmi(\alpha) \pi_{\eta_\rmi}^\mathcal{J}(\alpha)=
\eta_\rmi.
\end{equation}
\end{subequations}

\subsection{The new bounds}

\subsubsection{Hall's bound}

Let us consider now Holevo's bound for the new set up:
\begin{equation}
I_\rmc\{p_\rmf,\sigma;E_\mathcal{J}\} \leq \chi\{p_\rmf,\sigma\}.
\end{equation}
By \eqref{Iinvariance}, \eqref{rho2sigma} we get
\begin{equation}\label{Halldualbound}
I_\rmc\{p_\rmi,\rho_\rmi;E_\mathcal{I}\} \leq \chi\{p_\rmf,\sigma\}\equiv \sum_\omega
p_\rmf(\omega)\, S_\rmq\big(\sigma(\omega)\big\| \eta_\rmi\big),
\end{equation}
which is  Hall's bound $\big($Eq.\ (19) of \cite{Hal97}$\big)$. This bound is discussed also
in Refs.\ \cite{Hal97b,KinR01,Rus02}; the ``continuous'' version of it is given in
\cite{BarL04preB}.

\subsubsection{The new upper bound for $I_\rmc$}

Having defined a new instrument and not only a POV measure, we obtain from (\ref{SWW}) the
inequality
\begin{equation}\label{newub1}
\chi\{p_{\rmf},\sigma\}\geq I_\rmc\{p_\rmi,\rho_\rmi;E_\mathcal{I}\} + \sum_\alpha p_\rmi(
\alpha)\, \chi\big\{p_{\rmf|\rmi}(\bullet|\alpha), \pi_{\sigma(\bullet)}^\mathcal{J}
(\alpha)\big\},
\end{equation}
which gives a stronger bound than Hall's one (\ref{Halldualbound}). In order to render more
explicit this bound, it is convenient to start from the equivalent form (\ref{Iqineq}), which
now reads
\begin{equation}\label{nnn}
I_\rmq\{\eta_\rmi;\mathcal{J}\} \geq I_\rmc\{p_\rmi,\rho_\rmi;E_\mathcal{I}\} + \sum_\omega
p_\rmf(\omega)\, I_\rmq\{\sigma(\omega);\mathcal{J}\}.
\end{equation}
By Eqs.\ (\ref{newE}),   (\ref{newIq}),  \eqref{apostJ1} we obtain
\begin{equation}
I_\rmq\{\eta_\rmi;\mathcal{J}\} =\chi\{p_\rmi,\rho_\rmi\}.
\end{equation}
Therefore, Eq.\ (\ref{nnn}) gives the new bound
\begin{equation}\label{newbound}
I_\rmc\{p_\rmi,\rho_\rmi;E_\mathcal{I}\} \leq  \chi\{p_\rmi,\rho_\rmi\} - \sum_\omega p_\rmf
(\omega)\, I_\rmq\{\sigma(\omega);\mathcal{J}\};
\end{equation}
let us stress that $I_\rmq\{\sigma(\omega);\mathcal{J}\}\geq 0$ because of Eq.\ (\ref{newIq}).
More explicitly, by Eqs.\ (\ref{newE}), (\ref{newinistat}), (\ref{newIq}), we have
\begin{equation}
\sum_\omega p_\rmf( \omega)\, I_\rmq\{\sigma(\omega);\mathcal{J}\}= \sum_\omega p_\rmf(\omega)
\, S_\rmq\big(\sigma(\omega)\big)- \sum_{\alpha, \omega} p_{\rmi\rmf} ( \alpha, \omega)\,
S_\rmq \big(  \pi_{\sigma(\omega)}^\mathcal{J}(\alpha)\big),
\end{equation}
where $\sigma(\omega)$ is given by (\ref{newinistat}) and
$\pi_{\sigma(\omega)}^\mathcal{J}(\alpha)$ by \eqref{apostJ1}.  The general version of the
bound \eqref{newbound} has been presented in \cite{BarL04preB}.

\subsubsection{Scutaru's lower bound}

By \eqref{rho2sigma} one gets that the states $\xi$ \eqref{statesxi} have to be replaced by
\begin{equation}\label{epsScu}
\epsilon(\omega):= \sum_\alpha p_{\rmi|\rmf}(\alpha|\omega) \rho_\rmi(\alpha)\,;
\end{equation}
recalling also that $p_\rmi$ has to be replaced by $p_\rmf$, one gets that the bound
\eqref{lowbound1} becomes
\begin{equation}\label{scu2}
I_\rmc\{p_\rmi, \rho_\rmi;E_{\mathcal{I}}\} \geq \chi\{p_{\rmf},\epsilon\}\,.
\end{equation}
Note that
\begin{equation}
\sum_\omega p_\rmf(\omega)\epsilon(\omega)= \eta_\rmi\,.
\end{equation}
This bound was obtained, directly in the ``continuous case'', by Scutaru in \cite{Scu95}; he
used Uhlmann's monotonicity theorem and a ``classical$\to$quantum'' channel $\Psi$  mapping
states on $\mathcal{C}(A)$ (discrete probability densities on $A$) into states on $M_d$: if
$h$ is any discrete probability density on $A$, then
\begin{equation}
\Psi[h]= \sum_\alpha h(\alpha) \rho_\rmi(\alpha)\,.
\end{equation}
This channel is exactly the one we have used; indeed, with the symbols of Paragraph
\ref{para:newbound}, one can check that $\Psi= \Lambda_{E_\mathcal{J}}^{\eta_\rmi}$.
Therefore, Scutaru's channel $\Psi$ is the $\eta_\rmi$-transpose of the
``quantum$\to$classical'' channel associated to the POV measure introduced by Hall and Hall's
\eqref{Halldualbound} and Scutaru's \eqref{scu2} bounds are linked one to the other exctly as
Holevo's bound \eqref{Holb} is linked to the bound \eqref{lowbound1}.

\subsubsection{An upper bound on Holevo's $\chi$-quantity}

By \eqref{Iinvariance}, \eqref{rho2sigma}, \eqref{apostJ1}, inequality \eqref{varineq} gives
\begin{equation}\label{lastineq}
I_\rmc\{p_\rmi, \rho_\rmi;E_{\mathcal{I}}\}  +\sum_{\omega} p_{\rmf}(\omega) \,
\chi\{p_{\rmi|\rmf}(\bullet|\omega),\pi^{\mathcal{J}}_{\sigma(\omega)}\} \geq
\chi\{p_{\rmi},\rho_\rmi\};
\end{equation}
the average state of the ensemble
$\{p_{\rmi|\rmf}(\bullet|\omega),\pi^{\mathcal{J}}_{\sigma(\omega)}\}$ is
$\eta^\omega_\mathcal{J}$ defined in \eqref{newapr}. Let us stress that Holevo's
$\chi$-quantity depends only on the initial ensemble, while the l.h.s.\ of inequality
\eqref{lastineq} depends also on the POV measure.

In the Subsection \ref{subsec:mutent} all the inequalities of Section \ref{sec:Hol} have been
shown to be inequalities between mutual entropies. As the results of this section have been
obtained from those of Section \ref{sec:Hol} only by changing instrument, also all
inequalities of the present section can be obviously stated as inequalities between mutual
entropies.

\section{Summary of the inequalities and examples\label{sec:example}}

\subsection{The main inequalities}

The mutual information $I_\rmc\{p_\rmi,\rho_\rmi;E_\mathcal{I}\}$ is a key object, which
quantifies the ability of the POV measure $E_\mathcal{I}$ in extracting the information
codified in the initial ensemble. Let us summarize all the inequality involving
$I_\rmc\{p_\rmi,\rho_\rmi;E_\mathcal{I}\}$.

In Section \ref{sec:Hol} we obtained the new lower bound \eqref{lowbound1}, the generalization
\eqref{SWW} of the bound of Shumacher, Westmoreland, Wootters and Holevo's bound \eqref{Holb};
we can summarize their definitions and relationships by
\begin{subequations}
\begin{equation}\label{bnlb}
B_{\mathrm{Hlv}}:=\chi\{p_{\rmi},\rho_\rmi\}, \qquad b_{\mathrm{nlb}}:=\chi\{p_\rmi ,\xi\},
\end{equation}
\begin{equation}
\label{B_SWW} B_{\mathrm{SWW}} := \chi\{p_{\rmi},\rho_\rmi\} - \sum_{\omega} p_{\rmf}(\omega)
\, \chi\{p_{\rmi|\rmf}(\bullet|\omega),\rho_\rmf^\bullet(\omega)\},
\end{equation}
\end{subequations}
\begin{equation}
0\leq b_{\mathrm{nlb}} \leq I_\rmc\{p_\rmi,\rho_\rmi;E_\mathcal{I}\} \leq B_{\mathrm{SWW}}
\leq B_{\mathrm{Hlv}}\,.
\end{equation}
We are using $b$ for a lower bound and $B$ for an upper bound.

In Section \ref{sec:Hall} we obtained Scutaru's bound \eqref{scu2}, the new upper bound
\eqref{newbound} and Hall's bound \eqref{Halldualbound}; summarizing we have
\begin{equation}
0\leq b_{\mathrm{Scu}} \leq I_\rmc\{p_\rmi,\rho_\rmi;E_\mathcal{I}\} \leq
B_{\mathrm{nub}} \leq \begin{cases} B_{\mathrm{Hall}}
\\
B_{\mathrm{Hlv}}\end{cases}
\end{equation}
\begin{subequations}
\begin{gather}\label{bScu}
b_{\mathrm{Scu}}:=\chi\{p_{\rmf},\epsilon\},
\\
B_{\mathrm{nub}}:= \chi\{p_\rmi,\rho_\rmi\} - \sum_\omega p_\rmf (\omega) \,
I_\rmq\{\sigma(\omega);\mathcal{J}\},
\\
B_{\mathrm{Hall}}:=\chi\{p_\rmf,\sigma\}.
\end{gather}
\end{subequations}

Finally, the inequalities \eqref{aprioribound} and \eqref{lastineq} can be written as
\begin{equation}\label{2lb}
I_\rmc\{p_\rmi, \rho_\rmi;E_{\mathcal{I}}\}\geq \begin{cases} b_1
\\
b_2
\end{cases}
\end{equation}
\begin{subequations}
\begin{gather}\label{b_1}
 b_1:=
\chi\{p_{\rmi},\eta_\rmf^\bullet\}-\sum_{\omega} p_{\rmf}(\omega) \,
\chi\{p_{\rmi|\rmf}(\bullet|\omega),\rho^\bullet_\rmf(\omega)\} ,
\\ \label{b_2}
b_2 := \chi\{p_{\rmi},\rho_\rmi\}- \sum_{\omega} p_{\rmf}(\omega) \,
\chi\{p_{\rmi|\rmf}(\bullet|\omega),\pi^{\mathcal{J}}_{\sigma(\omega)}\} .
\end{gather}
\end{subequations}
However, $b_1$ and $b_2$ are not necessarily non-negative and, therefore, \eqref{2lb} does not
give always effective lower bounds on $I_\rmc$.

A notion related to that of classical mutual information, but not linked to a specific
measurement, is the accessible information of an ensemble \cite{Sch90}: it is the supremum
over all the POV measures of the classical mutual information extracted by the quantum
measurement
\begin{equation}
I_{\rm acc}\{p_\rmi,\rho_\rmi\}:= \sup_{E} I_\rmc\{p_\rmi,\rho_\rmi;E\}.
\end{equation}

The only bound from above for $I_{\rm acc}\{p_\rmi,\rho_\rmi\}$ is Holevo's one, because only
this bound does not depend on the measurement. From below $I_{\rm acc}\{p_\rmi,\rho_\rmi\}$ is
bounded by the subentropy introduced in \cite{JosRW94} and, trivially, by
$I_\rmc\{p_\rmi,\rho_\rmi;E\}$ computed for any fixed $E$ and by any of its lower bounds.

The subentropy of a density matrix $\rho$ is
\begin{equation}
Q(\rho)= - \sum_k \left(\prod_{\ell: {}\, \ell\neq k} \frac{\lambda_k}
{\lambda_k-\lambda_\ell} \right) \lambda_k \log \lambda_k\,,
\end{equation}
where the $\lambda_k$ are the eigenvalues of $\rho$ $\big($\cite{JosRW94}, Eq.~(8)$\big)$. The
bound based on the subentropy $\big($\cite{JosRW94}, Eq.~(33)$\big)$ is
\begin{equation}
I_{\rm acc}\{p_\rmi,\rho_\rmi\} \geq b_{\mathrm{subent}}\equiv Q(\eta_\rmi) -\sum_\alpha
p_\rmi(\alpha) Q\big(\rho_\rmi(\alpha)\big).
\end{equation}

\subsection{A rank-one POV measure}

As a first example, let us consider a measurement described by a POV measure made up of
rank-one elements:
\begin{subequations}\label{rank1}
\begin{gather}
E_\mathcal{I}(\omega)= \mu(\omega) |\psi(\omega)\rangle \langle \psi(\omega)|, \\ \|
\psi(\omega)\| =1\,, \qquad \mu(\omega)\geq 0\,, \qquad \sum_\omega \mu(\omega)
|\psi(\omega)\rangle \langle \psi(\omega)|=\openone\,.
\end{gather}
This gives
\begin{equation}
p_\rho(\omega)= \mu(\omega) \langle \psi(\omega)|\rho\,\psi(\omega)\rangle, \qquad \rho\in
\mathcal{S}_d\,,
\end{equation}
\end{subequations}
\begin{equation}
p_{\rmf|\rmi}(\omega|\alpha) = \mu(\omega) \langle
\psi(\omega)|\rho_\rmi(\alpha)\,\psi(\omega)\rangle, \qquad p_{\rmf}(\omega) = \mu(\omega)
\langle \psi(\omega)|\eta_\rmi\,\psi(\omega)\rangle,
\end{equation}
\begin{equation}
I_\rmc\{p_\rmi, \rho_\rmi;E_{\mathcal{I}}\}=\sum_{\alpha,\omega} p_\rmi(\alpha)\langle
\psi(\omega)|\rho_\rmi(\alpha)\,\psi(\omega)\rangle \mu(\omega)\, \log \frac{\langle
\psi(\omega)|\rho_\rmi(\alpha)\,\psi(\omega)\rangle}{\langle
\psi(\omega)|\eta_\rmi\,\psi(\omega)\rangle}\,.
\end{equation}

By \eqref{E&V} and the positivity of $\sum_{k\in K} V_k^{\omega \dagger}V_k^\omega$ one can
prove that for any instrument $\mathcal{I}$ compatible with the POV measure \eqref{rank1} it
must be
\begin{equation}
V_k^\omega=|\phi_k(\omega)\rangle \langle \psi(\omega)|, \qquad \sum_k \|
\phi_k(\omega)\|^2=\mu(\omega).
\end{equation}
By inserting this into the definition \eqref{apostI} of the a posteriori states, one gets that
\begin{equation}
\pi_\rho^\mathcal{I}(\omega)= \frac 1 {\mu(\omega)} \sum_k |\phi_k(\omega)\rangle
\langle\phi_k (\omega)|=: \pi(\omega),  \qquad \forall \rho\in \mathcal{S}_d;
\end{equation}
the a posteriori states depend on the instrument, but are independent from the
pre-measure\-ment state.

Then, we have $\rho^\alpha_\rmf(\omega)=\rho_\rmf(\omega)=\pi(\omega)$ and
\begin{equation}\label{vanishSWW}
\sum_{\omega} p_{\rmf}(\omega) \,
\chi\{p_{\rmi|\rmf}(\bullet|\omega),\rho_\rmf^\bullet(\omega)\}=0\,.
\end{equation}
Moreover, one can check that the states $\sigma(\omega)$ and
$\pi_{\sigma(\omega)}^\mathcal{J}(\alpha)$ are pure, that implies
\begin{equation}
\sum_\omega p_\rmf (\omega)\, I_\rmq\{\sigma(\omega);\mathcal{J}\}=0\,.
\end{equation}
The consequence is that the SWW bound \eqref{SWW} and the new upper bound \eqref{newbound}
reduce to Holevo's one \eqref{Holb}. Moreover, we get $ \chi\{p_\rmf,\sigma\}
=S_\rmq(\eta_\rmi)$; so, the original Hall's bound \eqref{Halldualbound} is worst than
Holevo's one, as already noticed by Hall himself \cite{Hal97}. Summarizing, the four upper
bounds are related by
\begin{equation}
B_{\mathrm{SWW}} =B_{\mathrm{nub}} = B_{\mathrm{Hlv}}\equiv  S_\rmq(\eta_\rmi)-
\sum_{\alpha\in A} p_{\rmi}(\alpha) S_\rmq(\rho_\rmi(\alpha))\leq B_{\mathrm{Hall}}\equiv
S_\rmq(\eta_\rmi).
\end{equation}

Let us consider now the lower bounds. The statistical operators $\xi$ and $\epsilon$ in the
new lower bound \eqref{bnlb} and in Scutaru's bound \eqref{bScu} are now given by
\begin{subequations}
\begin{gather}
\xi(\alpha)= \sum_\omega \mu(\omega)\,  \frac{\langle
\psi(\omega)|\rho_\rmi(\alpha)\,\psi(\omega)\rangle}{\langle
\psi(\omega)|\eta_\rmi\,\psi(\omega)\rangle} \, \eta_\rmi^{1/2} |\psi(\omega)\rangle \langle
\psi(\omega)| \eta_\rmi^{1/2}\,,
\\
\epsilon(\omega) = \sum_\alpha p_\rmi(\alpha)\,  \frac{\langle
\psi(\omega)|\rho_\rmi(\alpha)\,\psi(\omega)\rangle}{\langle
\psi(\omega)|\eta_\rmi\,\psi(\omega)\rangle}\, \rho_\rmi(\alpha).
\end{gather}
\end{subequations}
By \eqref{vanishSWW}, Eq.\ \eqref{b_1} gives the effective lower bound
\begin{equation}
b_1= \chi\{p_{\rmi},\eta_\rmf^\bullet\} \geq 0\,;
\end{equation}
moreover, the states $\eta^\alpha_\rmf$ turn out to be given by
\begin{equation}
\eta_\rmf^\alpha = \sum_{\omega} p_{\rmf|\rmi}(\omega|\alpha)\pi(\omega).
\end{equation}
Finally, by the fact that the states $\pi^{\mathcal{J}}_{\sigma(\omega)}(\alpha)$ are pure, we
get from \eqref{b_2}
\begin{equation}\label{lastineqvar1}
b_2=  \chi\{p_{\rmi},\rho_\rmi\}  -\sum_{\omega} p_{\rmf}(\omega) \, S_\rmq
(\eta_{\mathcal{J}}^{\omega}),
\end{equation}
with
\begin{equation}
\eta_{\mathcal{J}}^{\omega}= \frac 1 {\langle \psi(\omega)|\eta_\rmi\,\psi(\omega)\rangle}
\sum_\alpha p_\rmi(\alpha ) \rho_\rmi(\alpha)^{1/2} |\psi(\omega)\rangle \langle \psi(\omega)
| \rho_\rmi(\alpha)^{1/2}\,.
\end{equation}

\subsubsection{A complete von Neumann measurement\label{vNm}}

An interesting case of rank-one POV measure is certainly that one of a complete von Neumann
measurement. Let us consider here only the case of a projection valued measure, which
diagonalizes $\eta_\rmi$:
\begin{subequations}
\begin{gather}
\Omega=\{1,\ldots,d\}, \qquad  \langle \psi(\omega)|\psi(\omega^\prime)\rangle =
\delta_{\omega\omega^\prime}\,, \qquad \mu(\omega)= 1\,, \\ \eta_\rmi =\sum_{\omega =1}^d
\lambda_\omega |\psi(\omega)\rangle \langle \psi(\omega)|.
\end{gather}
\end{subequations}
Moreover, we construct the instrument by the usual reduction postulate, so that
\begin{equation}
\pi(\omega) =  E_\mathcal{I}(\omega) =|\psi(\omega)\rangle \langle \psi(\omega)|.
\end{equation}
Then, we have
\begin{gather}
p_{\rmf|\rmi}(\omega|\alpha) =  \langle \psi(\omega)|\rho_\rmi(\alpha)\,\psi(\omega)\rangle,
\qquad p_{\rmf}(\omega) = \lambda_\omega,
\\
I_\rmc\{p_\rmi, \rho_\rmi;E_{\mathcal{I}}\}=S_\rmq(\eta_\rmi) - \sum_{\alpha}
p_\rmi(\alpha)S_\rmc\big(p_{\rmf|\rmi}(\bullet|\alpha)\big).
\end{gather}
As before, only Holevo's bound survives as upper bound.

About the lower bounds, now we have
\begin{gather}
\epsilon(\omega) = \sum_\alpha p_\rmi(\alpha)\,  \frac{\langle
\psi(\omega)|\rho_\rmi(\alpha)\,\psi(\omega)\rangle}{\lambda_\omega}\, \rho_\rmi(\alpha),
\\
\eta_\rmf^\alpha = \xi(\alpha)= \sum_{\omega} \langle
\psi(\omega)|\rho_\rmi(\alpha)\,\psi(\omega)\rangle\pi(\omega).
\end{gather}
This gives
\begin{equation}
b_1= b_{\mathrm{nlb}}= I_\rmc\{p_\rmi, \rho_\rmi;E_{\mathcal{I}}\}\geq b_{\mathrm{Scu}}\equiv
S_\rmq(\eta_\rmi)-\sum_\omega \lambda_\omega S_\rmq\big(\epsilon(\omega)\big).
\end{equation}
Finally, $\eta_{\mathcal{J}}^{\omega}$ in $b_2$ \eqref{lastineqvar1} becomes
\begin{equation}
\eta_{\mathcal{J}}^{\omega}=\sum_\alpha  \frac {p_\rmi(\alpha )} {\lambda_\omega} \,
\rho_\rmi(\alpha)^{1/2} |\psi(\omega)\rangle \langle \psi(\omega) | \rho_\rmi(\alpha)^{1/2}\,.
\end{equation}

\subsubsection{The case of commuting letter states \label{commls}}

Let us consider now the case in which all the $\rho_\rmi(\alpha)$ are commuting operators; it
is known that this is the only case in which Holevo's bound is attained \cite{Hol73,Rus02}.

Let us choose $E_\mathcal{I}(\omega) =|\psi(\omega)\rangle \langle \psi(\omega)|$ to be a
joint spectral measure of all the operators $\rho_\rmi(\alpha)$; because, necessarily, also
$\eta_\rmi$ is diagonalized by $E_\mathcal{I}$, this is a particularization of the case of
Subsection \ref{vNm}. Then, we have
\begin{subequations}
\begin{gather}
\rho_\rmi(\alpha) =\sum_\omega \kappa_\omega^\alpha \pi(\omega), \qquad
\kappa_\omega^\alpha\geq 0\,, \quad \sum_\omega \kappa_\omega^\alpha =1\,, \quad \sum_\alpha
p_\rmi(\alpha)\kappa_\omega^\alpha = \lambda_\omega\,,
\\
\eta_{\mathcal{J}}^{\omega}=\pi(\omega), \qquad S_\rmq(\eta_{\mathcal{J}}^{\omega})=0\,,
\\
\epsilon(\omega) =\sum_{\omega^\prime} \frac{q_{12}(\omega,\omega^\prime)}{\lambda(\omega)}\,
\pi(\omega^\prime)\,, \qquad q_{12}(\omega,\omega^\prime):= \sum_\alpha
p_\rmi(\alpha)\kappa_{\omega}^\alpha \kappa_{\omega^\prime}^\alpha\,;
\end{gather}
\end{subequations}
let us note that $q_{12}$ is a joint discrete probability density with
marginals $q_{1}(\omega)=q_{2}(\omega)=\lambda_\omega$. Then, all the previous
equalities/inequalities reduce to
\begin{equation}
B_{\mathrm{Hall}} \geq B_{\mathrm{SWW}} = B_{\mathrm{nub}} =B_{\mathrm{Hlv}}= I_\rmc\{p_\rmi,
\rho_\rmi;E_{\mathcal{I}}\} = b_1 = b_2= b_{\mathrm{nlb}}\geq b_{\mathrm{Scu}} \equiv
S_\rmc(q_{12}\| q_1\otimes q_2).
\end{equation}

\subsection{Pure initial states\label{pureinst}}

When all the initial states $\rho_\rmi(\alpha)$ are pure, Holevo's $\chi$-quantity reduces to
the von Neumann entropy: $\chi\{p_\rmi,\rho_\rmi\}= S_\rmq(\eta_\rmi)$. Moreover, from Eqs.\
\eqref{newE}, \eqref{apostalpha} we have that $E_\mathcal{J}(\alpha)$ is a rank-one POV
measure and that $\mathcal{J}$ purifies any initial state: $\pi_\rho^\mathcal{J}(\alpha)=
\rho_\rmi(\alpha)$, $\forall \rho \in \mathcal{S}_d$. Then, Eqs.\ \eqref{79's}, \eqref{epsScu}
give
\begin{equation}
\pi_{\sigma(\omega)}^\mathcal{J}(\alpha)= \pi_{\eta_\rmi}^\mathcal{J}(\alpha)=
\rho_\rmi(\alpha), \qquad \eta_\mathcal{J}^\omega = \epsilon(\omega),
\end{equation}
which imply also
\begin{equation}
\sum_\alpha p_\rmi( \alpha)\, \chi\big\{p_{\rmf|\rmi}(\bullet|\alpha),
\pi_{\sigma(\bullet)}^\mathcal{J} (\alpha)\big\}=0\,.
\end{equation}
Therefore one obtains that inequality \eqref{newub1} reduces to Eq.\ \eqref{Halldualbound},
that Hall's bound is better than Holevo's bound in this case and that inequality
\eqref{lastineq} becomes equivalent to Scutaru's bound \eqref{scu2}:
\begin{equation}
b_2=b_{\mathrm{Scu}} \leq I_\rmc\{p_\rmi, \rho_\rmi;E_{\mathcal{I}}\} \leq B_{\mathrm{Hall}}
=B_{\mathrm{nub}} \leq B_{\mathrm{Hlv}}\equiv S_\rmq(\eta_\rmi).
\end{equation}

\subsubsection*{The instrument $\mathcal{I}$ is pure}

When the initial states are pure and, moreover, the instrument $\mathcal{I}$ sends pure states
into pure a posteriori states, one has also that the states $\rho_\rmf^\alpha(\omega)$ are
pure and
\[
\sum_{\omega} p_{\rmf}(\omega) \,
\chi\{p_{\rmi|\rmf}(\bullet|\omega),\rho_\rmf^\bullet(\omega)\}= \sum_{\omega}
p_{\rmf}(\omega) \,S_\rmq\big(\rho_\rmf(\omega)\big).
\]
Then, the SWW bound \eqref{B_SWW} reduces to
\begin{equation}\label{B_SWW2}
B_{\mathrm{SWW}} =I_\rmq(\eta_\rmi;\mathcal{I}) \equiv S_\rmq(\eta_\rmi) - \sum_\omega
p_{\rmf}( \omega) S_\rmq\big(\rho_\rmf(\omega)\big)\, .
\end{equation}

\subsection{Examples based on a two-level atom}

Here we give two examples based on a two-state system. This  case is particularly suited to
construct examples which allow for explicit calculations. The eigenvalues of a density matrix
$\rho\in \mathcal{S}_2$ are
\begin{subequations}
\begin{equation}
\lambda _\pm = \frac 1 2 \left( 1\pm \sqrt{1-4D}\right), \qquad D:=\det \rho\,, \qquad 0\leq D
\leq \frac 14\,.
\end{equation}
Then, the von Neumann entropy and the subentropy can be written as
\begin{gather}
S_\rmq(\rho)=\sqrt{1-4D} \left[ 1-\log \left(2\lambda_+\right)\right] - \lambda_-\log D\,,
\\
Q(\rho)= S_\rmq(\rho) - \frac{D}{\sqrt{1-4D}}\,\log \frac{\lambda_+}{ \lambda_-} \,.
\end{gather}
\end{subequations}

\subsubsection{Pure initial states and good counting measurement \label{goodmeas}}

Let us give now a simple example of the situation of Section \ref{pureinst}. We consider a
two-level atom whose ground and excited states are $|0\rangle=\binom{0}{1}$ and
$|1\rangle=\binom{1}{0}$, respectively. After the preparation, the atom is left isolated and,
if it is in the excited state, it can emit a photon. For what concerns the measurement, assume
that we are able only to count the number (0 or 1) of photons emitted in the time interval
$(0,t)$. The instrument is
\begin{subequations}
\begin{gather}
\mathcal{O}_t(0)[\rho]=\rme^{-\frac \Gamma 2 |1\rangle \langle 1|t}\, \rho\,\rme^{-\frac
\Gamma 2 |1\rangle \langle 1|t},
\\
\mathcal{O}_t(1)[\rho]=\int_0^t \rmd s\,\Gamma |0\rangle \langle
1|\mathcal{O}_s(0)[\rho]|1\rangle\langle 0| = \left( 1-\rme^{-\Gamma t}\right)|0\rangle
\langle 1|\rho |1\rangle\langle 0|,
\end{gather}
\end{subequations}
where $\Gamma $ is the decay rate. The associated POV measure is
\begin{equation}
E_t(0)= \rme^{-\Gamma t} |1\rangle \langle 1| +|0\rangle \langle 0|, \qquad
E_t(1)=\left(1-\rme^{-\Gamma t}\right) |1\rangle \langle 1|.
\end{equation}
In this example, due to the presence of the time $t$, we shall use the subscript ``$t$''
instead of ``$\rmf$'' for the final quantities; we shall also write the various bounds as
functions of $\Gamma t =: x$.

Assume that we are able to prepare the atom in the ground state $|0\rangle$ and, by a suitable
pulse, in the state $ \frac 1 {\sqrt{2}} \left(|0\rangle+|1\rangle \right)$; so, our initial
states are
\begin{equation}
\rho_\rmi(0) = |0\rangle \langle 0|=\begin{pmatrix} 0&0 \\ 0&1 \end{pmatrix}, \qquad
\rho_\rmi(1) =\frac 1 {2} \left(|0\rangle+|1\rangle \right)\left( \langle 0|+\langle 1|\right)
=\frac 1 2 \begin{pmatrix}  1 &  1 \\
1& 1
\end{pmatrix}.
\end{equation}
Moreover, let us assume that the a priori probabilities are equal:
\begin{equation}
p_\rmi(0)= p_\rmi(1)=\frac 12\,.
\end{equation}
Then, the initial average state is
\begin{equation}\label{etai1}
\eta_\rmi = \frac 1 4\begin{pmatrix} 1&1 \\
1&3
\end{pmatrix}.
\end{equation}

\begin{figure}[h]
\centerline{\epsfig{file=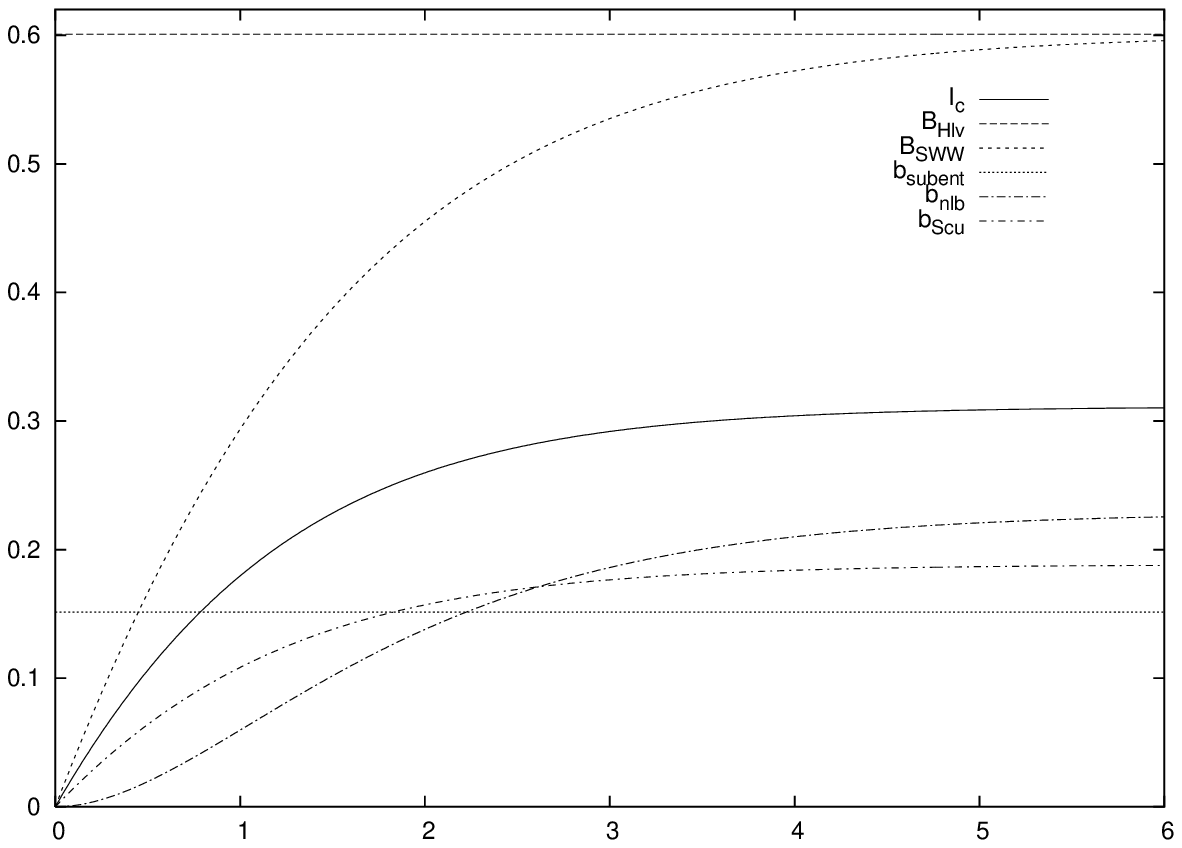}}

\bigskip
\fcaption{\label{fig1}The classical mutual information and the various bounds as functions of
$x=\Gamma t$: the example of Section \ref{goodmeas}. In this case
$B_{\mathrm{nub}}(x)=B_{\mathrm{Hall}}(x)= B_{\mathrm{SWW}}(x)$, $b_2(x)=b_{\mathrm{Scu}}(x)$,
$b_1(x)< 0$.}
\end{figure}

The various probability can be easily computed; we give the results in Appendix A. Then, the
explicit expression of the classical mutual information turns out to be
\begin{equation}
I_\rmc(\Gamma t):= I_\rmc\{p_\rmi, \rho_\rmi;E_{t}\} = \frac {3} 2+  \frac {1+\rme^{-\Gamma
t}} 4 \,\log \left(1+\rme^{-\Gamma t}\right)-\frac {3+\rme^{-\Gamma t}} {4} \,\log \left(3
+\rme^{-\Gamma t}\right);
\end{equation}
its maximum value is for large times:
\begin{equation}
\sup_{x>0} I_\rmc(x) =\lim_{x\to+\infty} I_\rmc(x) = \frac {3\left(2-\log 3\right)}4 \simeq
0.311278\,.
\end{equation}

Let us consider now the various bounds; all the determinants needed in the formulas are given
in Appendix A. First of all we have Holevo's bound and the subentropy bound
\begin{equation}
B_{\mathrm{Hlv}}= S_\rmq(\eta_\rmi) \simeq 0.600876\,, \qquad b_{\mathrm{subent}} =
Q(\eta_\rmi)= S_\rmq(\eta_\rmi)- \frac{\log \left(\sqrt{2}+1\right)} {2\sqrt{2}}\simeq
0.151314\,.
\end{equation}
The computations of the determinants give that also the SWW bound \eqref{B_SWW2} reduces to
Hall's one; we get
\begin{equation}
B_{\mathrm{nub}}(\Gamma t)=B_{\mathrm{Hall}}(\Gamma t)= B_{\mathrm{SWW}}(\Gamma t)=
S_\rmq(\eta_\rmi) - \frac{3+\rme^{-\Gamma t}}{4}\, S_\rmq\big(\rho_t (0)\big).
\end{equation}

Finally we have
\begin{equation}
b_{\mathrm{nlb}}(\Gamma t)=S_\rmq(\eta_\rmi) - \frac 12\, S_\rmq\big(\xi_t (0)\big)- \frac
12\, S_\rmq\big(\xi_t (1)\big),
\end{equation}
\begin{equation}
b_2(\Gamma t)=b_{\mathrm{Scu}}(\Gamma t)=S_\rmq(\eta_\rmi) - \frac {3+\rme^{-\Gamma t}}{4}\,
S_\rmq\big(\epsilon_t (0)\big).
\end{equation}

By numerical computations one can check that $b_1(\Gamma t) <0$. In Figure \ref{fig1} the
various bounds are plotted as functions of the length of the time interval $x= \Gamma t$.
%\begin{equation}
%b_1(\Gamma t) = S_\rmq(\eta_t)- \frac {1}{2} \, S_\rmq(\eta_t^1) -\frac {3+\rme^{-\Gamma t}}4
%\, S_\rmq\big(\rho_t (0)\big)\leq 0, \qquad \text{by numerical computations}
%\end{equation}

\subsubsection{Mixed initial states and imperfect measurement \label{badmeas}}

In the previous example many bounds turned out to be the same; to have a more generic
situation, we modify that example by rendering not pure one of the initial states and by
adding some more imperfection in the instrument.

We consider again a two-level atom, but now, when we try to count the number (0 or 1) of
photons emitted in the time interval $(0,t)$ a spurious count can be registered with a small
probability, due to some imperfection in the instrumentation. Let us say that now the
instrument is
\begin{subequations}
\begin{equation}
\mathcal{O}_t(1)[\rho]=\left( 1-\rme^{-\Gamma t}\right)\Big( \frac {49}{50}\, |0\rangle
\langle 1|\rho |1\rangle\langle 0| + \frac {1}{50}\,\rho\Big),
\end{equation}
\begin{equation}
\mathcal{O}_t(0)[\rho]=\frac {49}{50}\, \rme^{-\frac \Gamma 2 |1\rangle \langle 1|t}\,
\rho\,\rme^{-\frac \Gamma 2 |1\rangle \langle 1|t} + \frac{\rme^{-\Gamma t}} {50}\,\rho\,,
\end{equation}
\end{subequations}
where $\Gamma $ is the decay rate. The associated POV measure is
\begin{equation}
E_t(1)=\left(1-\rme^{-\Gamma t}\right) \Big(|1\rangle \langle 1| +\frac 1 {50}\, |0\rangle
\langle 0|\Big)\,, \qquad  E_t(0)= \rme^{-\Gamma t} |1\rangle \langle 1| +\frac{49+
\rme^{-\Gamma t}}{50}\,|0\rangle \langle 0|.
\end{equation}

We are able to prepare the atom in the ground state $|0\rangle$. We would also prepare the
state $ \frac 1 {\sqrt{2}} \left(|0\rangle+|1\rangle \right)$ by a suitable pulse, but some
imperfection again allows us only to obtain a mixture of this state with the ground state. So,
let us say that our initial states are
\begin{subequations}
\begin{gather}
\rho_\rmi(0) = |0\rangle \langle 0|=\begin{pmatrix} 0&0 \\ 0&1 \end{pmatrix},
\\
\rho_\rmi(1) =\frac 9 {10}\, \frac  12  \left(|0\rangle+|1\rangle \right)\left( \langle
0|+\langle 1|\right)+ \frac 1 {10}|0\rangle \langle 0|
=\begin{pmatrix}  9 /{20} &  9 /{20} \\
9 /{20}& {11} /{20}
\end{pmatrix}.
\end{gather}
\end{subequations}
Moreover, let us assume that the a priori probabilities are
\begin{equation}
p_\rmi(0)=\frac 4 9 \,, \qquad p_\rmi(1)=\frac 5 9\,.
\end{equation}
Then, the initial average state is the same as in the previous section:
\begin{equation}\label{etai2}
\eta_\rmi = \frac 1 4\begin{pmatrix} 1&1 \\
1&3
\end{pmatrix}.
\end{equation}

The various probabilities can be easily computed and are written down in Appendix B. Then, the
classical mutual information becomes
\begin{multline}
I_\rmc(\Gamma t):=I_\rmc\{p_\rmi, \rho_\rmi;E_{t}\} = \frac {1-\rme^{-\Gamma t}} {25}\Bigl(
\frac 2 9 \,\log \frac 4 {53}+\frac {461}{72}\,\log \frac {461}{265}\Bigr)
\\
{}+ \frac {98+2\rme^{-\Gamma t}} {225} \,\log \frac {4\left(49+\rme^{-\Gamma t}\right)}
{147+53\rme^{-\Gamma t}}+\frac {539+461\rme^{-\Gamma t}} {1800} \,\log
\frac{539+461\rme^{-\Gamma t}}{5\left(147+53\rme^{-\Gamma t}\right)} \\ \overset{t\to
\infty}{\simeq} 0.21822\,.
\end{multline}

\begin{figure} [h]
\centerline{\epsfig{file=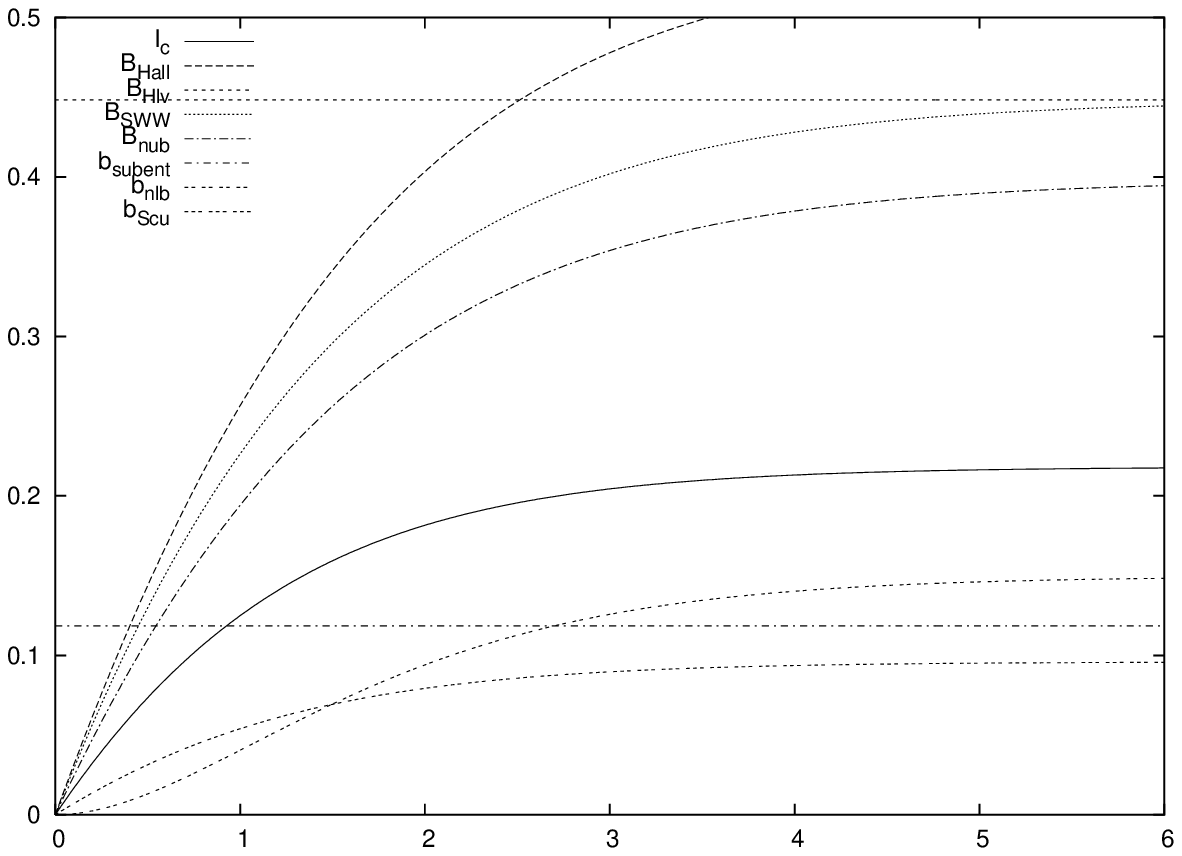}}

\bigskip \fcaption{\label{fig2}The classical mutual information and the various bounds as
functions of $x=\Gamma t$: the example of Section \ref{badmeas}. In this case $b_1(x) <0$,
$b_2(x) \leq 0$.}
\end{figure}

To calculate the various bounds, we need many determinants, again given in Appendix B. Then,
we have the various bounds: Holevo's bound
\begin{equation}
B_{\mathrm{Hlv}}:= \chi\{p_\rmi,\rho_\rmi\}= S_\rmq(\eta_\rmi)-\frac 5 9 \,
S_\rmq\big(\rho_\rmi(1)\big) \simeq 0.448368\,,
\end{equation}
Hall's bound
\begin{equation}
B_{\mathrm{Hall}}(\Gamma t):=\chi\{p_t,\sigma_t\} =S_\rmq(\eta_\rmi) - p_t (0)
S_\rmq\big(\sigma_t (0)\big)- p_t (1) S_\rmq\big(\sigma_t (1)\big),
\end{equation}
the new lower bound
\begin{equation}
b_{\mathrm{nlb}}(\Gamma t):= \chi\{p_\rmi,\xi_t\} =S_\rmq(\eta_\rmi) - \frac 4 9\,
S_\rmq\big(\xi_t (0)\big)- \frac 5 9\, S_\rmq\big(\xi_t (1)\big),
\end{equation}
Shumacher-Westmoreland-Wootters' bound
\begin{multline}
B_{\mathrm{SWW}}(\Gamma t):=\chi\{p_\rmi,\rho_\rmi\}-\sum_{\omega} p_{t }(\omega) \,
\chi\{p_{\rmi|t }(\bullet|\omega),\rho_t ^\bullet(\omega)\}
\\ {}
=B_{\mathrm{Hlv}}-\sum_{\omega}\left[ p_{t }(\omega)S_\rmq\big(\rho_t(\omega)\big) - p_{\rmi t
}(1,\omega)S_\rmq\big(\rho_t^1(\omega)\big)\right],
\end{multline}
the new upper bound
\begin{multline}
B_{\mathrm{nub}}(\Gamma t):= \chi\{p_\rmi,\rho_\rmi\}-\sum_\omega p_t(\omega)
I_\rmq\{\sigma_t(\omega);\mathcal{J}\} \\ {}= B_{\mathrm{Hall}}(\Gamma t)- \frac 5 9 \,
S_\rmq\big(\rho_\rmi(1)\big)+ \sum_\omega p_{\rmi
t}(1,\omega)S_\rmq\big(\pi^\mathcal{J}_{\sigma_t(\omega)}(1)\big),
\end{multline}
Scutaru's bound
\begin{equation}
b_{\mathrm{Scu}}(\Gamma t):=\chi\{p_t,\epsilon_t\}=S_\rmq(\eta_\rmi)
-p_t(0)S_\rmq\big(\epsilon_t(0)\big) -p_t(1)S_\rmq\big(\epsilon_t(1)\big),
\end{equation}
the subentropy lower bound for the accessible information
\begin{subequations}
\begin{gather}
b_{\mathrm{subent}}=B_{\mathrm{Hlv}}-d(0.125)+ \frac 5 9 \, d(0.045)\simeq 0.118467\,,
\\
d(x):= \frac x{\sqrt{1-4x}}\, \log \frac{1+\sqrt{1-4x}}{1-\sqrt{1-4x}}\,.
\end{gather}
\end{subequations}

%\begin{equation}
%%b_1(\Gamma t) = S_\rmq(\eta_t) - \frac 5 9 \, S_\rmq(\eta^1_t) +B_{\mathrm{SWW}}(\Gamma
%%t)-B_{\mathrm{Hlv}}
%%\\
%b_2(\Gamma t) =B_{\mathrm{Hlv}} - \sum_\omega p_t(\omega ) \left[ S_\rmq
%\left(\eta_{\mathcal{J}}^\omega\right) - p_{\rmi|t}(1|\omega)\,  S_\rmq
%\left(\pi^{\mathcal{J}}_{\sigma_t(\omega)}(1)\right)\right]
%\end{equation}

By numerical computations one can check that $b_1(\Gamma t) <0$ and $b_2(\Gamma t) \leq 0$. In
Figure \ref{fig2} the various bounds are plotted as functions of the length of the time
interval $x= \Gamma t$.

\subsubsection{A special feature of the two ensembles}

In Section \ref{vNm} we have considered a POV measure made up of the eigenprojections of the
initial average state $\eta_\rmi$ and in Section \ref{commls} we have recalled that this
choice saturates Holevo's inequality in the case of commuting letter states. However, when the
letter states do not commute, not only the eigenprojections of $\eta_\rmi$ do not give
necessarily the best measurement, but they can even be the worst choice, as shown by the case
of the ensembles of Sections \ref{goodmeas} and \ref{badmeas}.

The average state $\eta_\rmi$ is the same in both cases, see Eqs.\ \eqref{etai1} and
\eqref{etai2}. Its eigenprojections are $P_\pm = \frac 1 {2\sqrt{2}} \begin{pmatrix} \sqrt{2}
\mp 1 & \pm 1 \\ \pm 1 & \sqrt{2}\pm 1 \end{pmatrix}$, for which we get $\Tr \{P_\pm \rho\}=
\frac {2\pm \sqrt{2}} 4$ for any density matrix of the form $\rho= \begin{pmatrix} a & a
\\ a & 1-a \end{pmatrix}$. But this is the form of all the letter states of Sections
\ref{goodmeas} and \ref{badmeas}; therefore, in both cases,
$p_{\rmf|\rmi}(\pm|\alpha)=p_\rmf(\pm)$ and, so, $I_\rmc=0$.

\nonumsection{Acknowledgements}

\noindent Work supported by the \emph{European Community's Human Potential Programme} under
contract HPRN-CT-2002-00279, QP-Applications.

\nonumsection{References}

\appendix{. Two-level system, first example\label{app:A}}
The various probabilities needed in the example are
\begin{subequations}
\begin{gather}
p_{t |\rmi}(0|0) =1\,, \quad p_{t |\rmi}(1|0) =0\,, \qquad p_{t |\rmi}(0|1) =\frac  {
1+\rme^{-x}}{2} \, ,  \quad p_{t |\rmi}(1|1) = \frac { 1-\rme^{-x}}{2}\,,
\\
p_{t }(0) =\frac {3  +\rme^{-x}} 4\,, \qquad p_{t }(1) = \frac {1 -\rme^{-x}} 4\,,
\\
p_{\rmi t }(0,0) =\frac 1 2\,, \quad  p_{\rmi t }(1,0)=\frac { 1+\rme^{-x}}{4}\,, \quad
p_{\rmi t }(0,1)=0\,, \quad  p_{\rmi t }(1,1) = \frac { 1-\rme^{-x}} 4\,,
\\
p_{\rmi|t }(0|1) =0\,, \quad p_{\rmi|t }(1|1) =1\,, \qquad p_{\rmi|t }(0|0) =\frac {2} {
3+\rme^{-x}}\,, \quad p_{\rmi|t }(1|0) = \frac {1+\rme^{-x}} { 3+\rme^{-x}}\,.
\end{gather}
\end{subequations}

For what concerns the determinants involved in the upper bounds, we have
\begin{equation}
\det \eta_\rmi=\frac 1 8\,, \qquad \det \rho_\rmi(\alpha) = 0\,.
\end{equation}
Then, Eq.\ \eqref{newinistat} gives \ $\displaystyle \det \sigma_t(\omega)=\frac{\det
\eta_\rmi \;\det E_t(\omega)} {p_t(\omega)^2}$ \ and we get
\begin{equation}
\det \sigma_t(0) =\frac {2\,\rme^{-x}}{\left(3+\rme^{-x}\right)^2}\,, \qquad \det
\sigma_t(1)=0\,.
\end{equation}
By direct computations, we obtain
\begin{equation}
\det \rho_t ^\alpha(\omega)=\det \rho_t (1)=0\,, \qquad \det \rho_t (0)=\frac
{2\,\rme^{-x}}{\left(3+\rme^{-x}\right)^2}\,,
\end{equation}
\begin{equation}
\det \eta_t^0=0\,, \qquad \det \eta_t ^1 = \frac{\rme^{-x}}{4} \left(1-\rme^{-x}\right),
\qquad \det \eta_t  = \frac{\rme^{-x}}{16} \left(3-\rme^{-x}\right).
\end{equation}
Finally, we get
\begin{subequations}
\begin{equation}
\xi_t(0)= \sigma_t(0), \qquad \xi_t(1)= \frac 2 {3+\rme^{-x}}\, \eta_\rmi^{1/2} \left[
\left(3-\rme^{-x}\right)|1\rangle \langle 1 | + \left(1+\rme^{-x}\right)|0\rangle \langle 0
|\right]  \eta_\rmi^{1/2}\,,
\end{equation}
\begin{equation}
\det \xi_t(0)= \frac {2\,\rme^{-x}}{\left(3+\rme^{-x}\right)^2}\,,\qquad \det \xi_t(1)=
\frac{\left(3-\rme^{-x}\right) \left(1+\rme^{-x}\right)}{2\left(3+\rme^{-\Gamma
t}\right)^2}\,,
\end{equation}
\end{subequations}
\begin{subequations}
\begin{equation}
\epsilon_t(1) =\rho_\rmi(1), \qquad \epsilon_t(0)= \frac 1 {2\left(3+\rme^{-x}\right)}
\begin{pmatrix} 1+\rme^{-x} & 1+\rme^{-x} \\ 1+\rme^{-x} &
5+\rme^{-x} \end{pmatrix},
\end{equation}
\begin{equation}
\det \epsilon_t(0)= \frac {1+\rme^{-x}}{\left(3+\rme^{-x}\right)^2}\,, \qquad \det
\epsilon_t(1) =0\,.
\end{equation}
\end{subequations}

\appendix{. Two-level system, second example\label{app:B}}

First of all, the various probabilities are
\begin{subequations}
\begin{alignat}{2}
&p_{t |\rmi}(0|0) =\frac{49+\rme^{-x}} {50}\,, \qquad &p_{t |\rmi}(1|0) =\frac{1-\rme^{-x}}
{50}\,,
\\
&p_{t |\rmi}(0|1) =\frac  { 539+461\,\rme^{-x}}{1000} \, ,\qquad &p_{t |\rmi}(1|1) = \frac  {
461\left(1-\rme^{-x}\right)}{1000}\,,
\\
&p_{t }(0) =\frac {147  +53\,\rme^{-x}} {200}\,, \qquad &p_{t }(1) = \frac {53\left(1
-\rme^{-x}\right)} {200}\,,
\\
&p_{\rmi t }(0,0) =\frac {2\left(49  +\rme^{-x}\right)} {225}\,, \qquad  &p_{\rmi t
}(1,0)=\frac {539 +461\,\rme^{-x}} {1800}\,,
\\
&p_{\rmi t }(0,1)= \frac  { 2\left(1-\rme^{-x}\right)}{225}\,, \qquad  &p_{\rmi t }(1,1) =
\frac { 461\left(1-\rme^{-x}\right)}{1800}\,,
\\
&p_{\rmi|t }(0|1) =\frac{16}{477}\,, \qquad &p_{\rmi|t }(0|0) =\frac {16\left(49+\rme^{-x}
\right)} {9\left( 147+53\,\rme^{-x}\right)}\,,
\\
&p_{\rmi|t }(1|1) =\frac{461}{477}\,, \qquad &p_{\rmi|t }(1|0) = \frac {539+461\,\rme^{-x}}
{9\left( 147+53\,\rme^{-x}\right)}\,.
\end{alignat}

Then, Eqs.\ \eqref{newinistat}, \eqref{statesxi},  \eqref{apostJ1} give
\end{subequations}
\begin{equation}
\det \sigma_t(\omega)=\frac{\det \eta_\rmi \;\det E_t(\omega)} {p_t(\omega)^2}\,, \qquad \det
\pi_{\sigma_t(\omega)}^\mathcal{J}(\alpha)=\frac{\det \rho_\rmi(\alpha) \;\det E_t(\omega)}
{p_{t|\rmi}(\omega|\alpha)^2}\,,
\end{equation}
\begin{equation}
\det \xi_t(\alpha)=\det \eta_\rmi  \;\det \Big[\sum_\omega \frac{p_{t|\rmi}(\omega|\alpha)} {
p_{t}(\omega)}\, E_t(\omega)\Big].
\end{equation}

The final result of the computations of the determinants are
\begin{equation}
\det \eta_\rmi=\frac 1 8\,, \qquad \det \rho_\rmi(1) = \frac 9{200}\,, \qquad \det
\rho_\rmi(0) =\det \eta_t^0=0\,,
\end{equation}
\begin{equation}
\det \eta_t^1= \frac{9 \left[\left(1+49\, \rme^{-x}\right) \left(1991-441\, \rme^{-x}\right) -
9 \left(1+49\, \rme^{-x/2}\right)^2 \right]} {10^6}\,,
\end{equation}
\begin{equation}
\det \eta_t= \frac{\left(1+49\, \rme^{-x}\right) \left(199-49\, \rme^{-x}\right) -
\left(1+49\, \rme^{-x/2}\right)^2 } {4\times 10^4}\,,
\end{equation}
\begin{equation}
\det \sigma_t(0)=\frac{100\,\rme^{-x}\left(49+\rme^{-x}\right)} {\left(147+53\,\rme^{-x}
\right)^2}\,, \qquad \det \sigma_t(1)=\left(\frac {10}{53}\right)^2\,,
\end{equation}
\begin{equation}
\det \xi_t(0)=\frac {4\left(1274+51\,\rme^{-x}\right)\left(147+2503\,\rme^{-x }
\right)}{\left[53\left(147+53\,\rme^{-x}\right)\right]^2}\,,
\end{equation}
\begin{equation}
\det \xi_t(1)=\frac {\left(67767-14767\,\rme^{-x}\right)\left(29351+23649\,\rme^{-x}
\right)}{2 \left[ 530\left(147+53\,\rme^{-x}\right)\right]^2}\,,
\end{equation}
\begin{equation}
\det \rho_t ^0(\omega) =0, \qquad \det \rho_t ^1(1) =\frac{9\times 443}{(461)^2}\,, \qquad
\det \rho_t (1) =\frac{51}{(53)^2}\,,
\end{equation}
\begin{equation}
\det \rho_t^1(0)=\frac {9\,\rme^{-x}}{\left(539+461\,\rme^{-x}\right)^2}\, \left(
5341-882\,\rme^{-x/2} +541\,\rme^{-x}\right),
\end{equation}
\begin{equation}
\det \rho_t (0)=\frac {\rme^{-x}\left(4949+149\,\rme^{-x}- 98\,\rme^{-x/2}\right)
}{\left(147+53\,\rme^{-x}\right)^2}\,,
\end{equation}
\begin{equation}
\det \pi^\mathcal{J}_{\sigma_t(\omega)}(0)=0, \qquad \det \pi^\mathcal{J}_{\sigma_t(1)}(1)=
\left(\frac{30}{461}\right)^2,
\end{equation}
\begin{equation}
\det \pi^\mathcal{J}_{\sigma_t(0)}(1)= \frac{900\,\rme^{-x}\left(49+\rme^{-x}\right)}
{\left(539+461\,\rme^{-x}\right)^2}\,,
\end{equation}
\begin{equation}
\det \epsilon_t(0)= \frac{\left(539+461\,\rme^{-x}\right) \left(931+69\,\rme^{-x}\right)}
{200\left(147+53\,\rme^{-x}\right)^2}\,, \qquad \det  \epsilon_t(1)= \frac{ 69\times
461}{200\times 53^2}\,.
\end{equation}

\end{document}